\documentstyle{article}
\input boxedeps.tex
\SetRokickiEPSFSpecial  
\HideDisplacementBoxes
\newcommand{\sit}{\sigma_{T}}
\newcommand{\sil}{\sigma_{L}}
\newcommand{\D}{\overline{D}}
\newcommand{\da}{\dagger}  
\newcommand{\be}{\begin{equation}}
\newcommand{\eq}{\end{equation}}
   
\newcommand{\Tr}{{\rm \, Tr \!}}    
\newcommand{\integer}{\cal Z}       
\newcommand{\newl}{l}               
\newcommand{\newtau}{t}             
\newcommand{\firstpaper}{{\bf I}}   
\newcommand{\klink}{P^+_{\rm link}} 
\newcommand{\dklink}{K_{\rm link}}  
\newcommand{\kmax}{{P_{\rm max}}} 
\newcommand{\dkmax}{K_{\rm max}}    
\newcommand{\dm}{{\cal M}}          

\begin{document}

\mbox{}\hfill FAU--TP3--98/12\\
\mbox{}\hfill CERN--TH/98--232\\
\mbox{}\hfill {\tt hep-th/9806231}\\
\vspace{10mm}
\begin{center}
{\LARGE  Transverse Lattice Approach to \\ Light-Front
Hamiltonian QCD \\}
\vspace{30mm}

{\bf S. Dalley}${}^{*}$\footnote{On leave from DAMTP, University of Cambridge,
Cambridge CB3 0DS, England.} and  {\bf B. van de Sande}${}^{**}$\\
\vspace{10mm}

{\em
${}^*$Theory Division, CERN, CH--1211 Geneva 23, Switzerland
\vspace{5mm}

${}^{**}$Institut F\"ur Theoretische Physik III,\\
Staudstra{\ss}e 7, D--91058 Erlangen, Germany }
\vspace{30mm}

\end{center}

\begin{abstract}
We describe a non-perturbative  procedure for solving from first principles
the light-front Hamiltonian problem of $SU(N)$ pure gauge theory in
$D$ spacetime dimensions ($D \geq 3$),
based on enforcing Lorentz covariance of observables. A
transverse lattice regulator and colour-dielectric link fields are
employed,
together with an associated effective potential.
We argue that the light-front vacuum is necessarily trivial for large
enough lattice spacing, and clarify why this leads to an
Eguchi-Kawai dimensional reduction of observables to $1+1$-dimensions
in the $N\to\infty$ limit.
The procedure is then tested by explicit calculations for $2+1$-dimensional
$SU(\infty)$ gauge theory, within
a first approximation to the lattice effective potential.
We identify a scaling trajectory which  produces Lorentz covariant
behaviour for the lightest glueballs. 
The predicted masses, in units of the measured string tension,
are in agreement with recent results from conventional 
Euclidean lattice simulations. 
In addition, we obtain 
the  potential between heavy  sources and  the structure of the 
glueballs from their  light-front
wavefunctions.
Finally, we briefly discuss the extension of these 
calculations to $3+1$ dimensions.

\end{abstract}

\newpage
\baselineskip .25in


\section{Introduction}
Light-front Hamiltonian quantisation is a very promising approach to the
non-perturbative calculation and understanding 
of hadronic substructure, as a function of the quark and gluonic
degrees of freedom.
This is both of intrinsic interest
and important for the proper identification of new physics
beyond the Standard Model.
Recently, we have developed an idea \cite{us1}, originally suggested
by Bardeen and Pearson many years ago \cite{bard1,bard2},
for formulating the light-front Hamiltonian
of QCD on a transverse lattice.
The continuum limit is taken in time
and one space direction, while the remaining transverse degrees of freedom
are represented as colour-dielectric link variables \cite{mack} 
in a gauge invariant way on the transverse lattice.
Since light-front wavefunctions carry the  non-perturbative, coherent
information about hadronic bound and scattering states in a general 
Lorentz frame, a solution to this problem would open
up a vast amounts of existing and future experimental results to 
theoretical scrutiny \cite{brodsky}. We believe that the 
menagerie of light-front Hamiltonian, transverse lattice,
and effective colour-dielectric fields, is  a 
very potent combination in the analysis of the strong interactions at
intermediate scales. Each
contributes to the solution of long-standing technical problems associated with
the others. In addition, profound simplifications take place when one
considers only the leading order of the
$1/N$ expansion \cite{hoof} in the number of colours $N$. 
In this paper, we describe in detail
how non-perturbative first principles calculations may be performed
for pure gauge theory, in $2+1$ and $3+1$ dimensions,
by imposing Lorentz covariance on physical observables.
The procedure is then tested with explicit numerical and analytic
computations for $2+1$ dimensions at large $N$, 
comparing successfully with recent `state-of-the-art'
results
on the glueball spectrum from conventional Euclidean lattice simulations, 
as well as
producing many new results.

The choice of dynamical variables and regulator in canonical
light-front quantisation is crucial to a practical solution. 
The well-known Lorentz boost invariance of light-front 
wavefunctions comes with a price, because, for example, 
the light-front quantisation 
surface is not rotationally invariant; 
typically the regulators chosen
will break rotational invariance. An effective potential must be 
introduced to restore full Lorentz invariance. This effective
potential plays another important role to encode what is usually,
{\em id est} in equal-time quantisation, termed `vacuum structure.'
One of the attractive features of light-front quantisation is that,
in the true groundstate of a theory, the light-front vacuum is trivial
--- there can be no pair creation ---
except possibly for some massless gauge modes of zero momentum.
But in the true groundstate, much of the complication usually attributed 
to the vacuum has only been shifted into effective interactions appearing
in the Hamiltonian. 

Typically a light-front Hamiltonian will consist of a `bare' part
$H_B$ derived from the classical action, and the effective potential 
consisting of counterterms $H_C$. The role of the latter
is to enforce the fundamental (unbroken) symmetries required of the theory, 
which are broken by the choice of regulators. 
For practical purposes, one must decide on some physical grounds 
which subset of 
all possible operators one is going to consider in $H_C$, a subset which will 
then be systematically enlarged as greater accuracy is demanded.
The method we will use in our work, to fix the couplings to this
subset of operators, is to solve the entire theory 
as a function of the 
couplings, which are then tuned by demanding unbroken symmetries of
calculated observables. 
This straightforward approach will allow direct calculations
involving typical hadronic scales, intermediate
between the asymptotically free region and nuclear scales.
Naively, one might think that the
approach is computationally unviable, but this is not 
the case if one uses  Hamiltonian methods.
There already exists at least one successful application
of this approach to fixing counterterms. Parity, which is not a 
manifest symmetry of the light-front, is typically broken by
regulators. Burkardt \cite{burk} has shown how, 
in the case of the Yukawa model in $1+1$ dimensions,
a simple counterterm coupling can be adjusted to restore parity invariance
in a suitable observable. The general problem in higher dimensions 
involves the far more complicated task of ensuring the continuous 
Lorentz symmetries also.  This will form the main part of our job in solving
pure gauge theory.

To realise the above scenario, we apply light-front quantisation 
to a set of massive effective degrees of freedom on the transverse lattice
--- colour-dielectric link variables --- which efficiently organise
a gauge invariant Hamiltonian in the true vacuum. These variables are
supposed to represent an average over the high frequency transverse
gluon degrees of 
freedom. The fact that (lattice) gauge
invariance is not broken helps to restrict the number of 
counterterms needed in $H_C$, and leads to a simple understanding
of confinement.
The dielectric variables will provide the required physical basis for 
systematic approximation to the low-energy effective potential.
Moreover, they are flexible enough that in the light-front
Hamiltonian formulation
one can solve for  physical  
observables, with reasonable computational effort, 
as a function of the couplings in the effective
potential. This formulation therefore satisfies the requirements necessary
for our  approach. In the first half of this paper, we 
construct in detail the framework for making calculations in pure gauge
theory at large $N$. 
This framework is developed in $3+1$ dimensions, but can also
be applied in $2+1$ dimensions with some trivial modifications.

As a 
quantitative test in a highly relativistic,
strongly interacting system,  we then perform
detailed calculations in $2+1$ dimensions.
This choice as a testing ground is motivated by a number
of reasons.
A more reliable computation can be performed in $2+1$ dimensions than
$3+1$ dimensions.
Teper \cite{teper} has recently completed extensive and precise 
conventional Euclidean lattice Monte Carlo (ELMC) simulations
in three dimensions for various $N$.  The resulting spectra
may be extrapolated to large $N$, providing a benchmark for our work.
Finally, the physics of Yang-Mills theory in $2+1$
dimensions is remarkably similar that in $3+1$ dimensions.
%
In our previous work \cite{us1} (hereafter referred to as \firstpaper{}), 
we  studied  the  effective potential in $2+1$ dimensions
by matching our glueball spectrum to Teper's preliminary data.
For a truncated effective potential,
we were able to identify a coupling constant trajectory along
which mass ratios were approximately constant and agreed reasonably well
with the ELMC continuum predictions. 
This is a perfectly acceptable phenomenological approach, but one would
obviously  like to avoid fitting to an existing spectrum, 
whether taken from experiment
or another theoretical method. The first principles calculations we 
perform in this paper use only 
general concepts like gauge and Lorentz invariance to fix the
effective potential. In this way, we can make independent predictions 
of all physical quantities.
The contents  of the present paper are  summarised below.


In the next section, we briefly review the colour-dielectric
formulation of lattice gauge theory and its light-front Hamiltonian
limit in order to set notation. In Section~\ref{clcq}, we carry out
the light-front quantisation, paying particular attention to the
interplay of cutoffs, gauge fixing, and energetics. This allows us to
show that in the `colour-dielectric regime' where the light-front
Hamiltonian is quantised, light-front zero modes may be neglected,
{\em id est} the light-front vacuum is essentially trivial (this justifies the
naive light-front gauge 
choice often used).  We observe that linear confinement is
generic to the theory in the colour-dielectric regime.  All our work
is performed in the large-$N$ limit because of remarkable
simplifications which occur.  In Ref.~\cite{dalley1} and \firstpaper{}
it was proved that, in this limit, one effectively solves the
light-front Hamiltonian of a $1+1$-dimensional gauge theory coupled to
$2^{D-2}$ adjoint matter fields, for $D$ space-time  dimensions.
We clarify in Section~\ref{largen}
the physical role of this Eguchi-Kawai dimensional reduction. Unlike
lattice path integral quantisation, where in the leading order in $N$
vacuum structure undergoes dimensional reduction to zero dimensions,
the leading order in $N$ of the transverse lattice light-front
quantisation exhibits dimensional reduction of physical observables to
$1+1$ dimensions.  Section~\ref{tensionsection} outlines procedures
for measuring the string tension in the transverse lattice direction
via winding modes,
and the full spatial potential between two heavy sources, extending
work of Ref.~\cite{burkardt}. 
In Section~\ref{lorentzinvariance}, we construct states with nonzero 
transverse momentum.  We use the resulting dispersion relation,
together with the string tension measurements,
to construct a simple algorithm for fixing the effective
potential based on Lorentz covariance.

We then turn to a numerical implementation of our procedure for 
$2+1$-dimensional gauge theory (Section~\ref{calculation}). 
Most of the relevant formulae and
descriptions of our numerical methods have been given in 
\firstpaper{}; we review and extend them in Sections~\ref{twoplusone}, 
\ref{methods}, and the Appendices. Section~\ref{results} 
presents our results for the
scaling behaviour, glueball masses and wavefunctions, 
the string tensions and heavy-source potential.
Finally, we conclude with a brief discussion of the extension to $3+1$
dimensions of these calculations.


\section{Colour-Dielectric Lattice Gauge Theory.}

\subsection{Euclidean Lattice}
\label{euclid} 

For the purposes of studying a light-front Hamiltonian limit, Wilson's
formulation of lattice gauge theory \cite{wilson} leads to difficulties
due to the non-linearity of the link variables \cite{griff}.
We therefore take as our starting point Mack's linearised `colour-dielectric'
formulation of 
lattice gauge theory \cite{mack}.
%
%
A similar lattice gauge theory was first
introduced by Weingarten \cite{wein} as a generator of the spacetime lattice
regularisation of the Nambu string action.

On a 4-dimensional hypercubic lattice of spacing $a$, 
introduce a complex matrix 
$M_\mu(x) \in GL(N,C)$ for the link attached to site 
$x = \left\{x^0,x^1,x^2,x^3\right\}$
in the direction $x^\mu$. The partition function is
\be
        Z = \int \prod_{x,\mu} \prod_{i,j =1}^{N} 
        dM_{\mu,ij}(x) \, dM^*_{\mu,ij}(x) 
        \, \exp(- S[M]) \label{smear} \; ,
\eq
with colour indices $i,j \in\left\{1,\cdots,N\right\}$ and
$M_\mu(x) = M^{\da}_{-\mu}(x+a \widehat{\mu})$.
Throughout, $\widehat{\mu}$ is the unit vector in the $x^\mu$
direction and $\widehat{-\mu}=-\widehat{\mu}$.
For an appropriate choice of action $S$, this theory is 
invariant under lattice gauge transformations
\be
M_\mu(x) \to U^{\dagger}(x) M_\mu(x)  
U(x + a\widehat{\mu})  \; , \; \; \;\;  U \in SU(N) \;.
\eq
Since we will eventually be interested in 
an expansion of the action in powers
of the field $M$ in the large $N$ limit, we take $S[M]$ to consist
of  Wilson loops and their products. 
Note that this will include Wilson lines
which `backtrack', such as $\Tr\left\{M_\mu(x) M_\mu^{\da}(x) \right\}$,
 because $M$ is not necessarily a unitary matrix. (Determinants of $M$ are
also gauge invariant, but involve an infinite product of fields when 
$N=\infty$.)
Since we are interested in QCD,  $S$ is tuned in such a
way that, as the correlation length approaches infinity in direction $x^\mu$
(lattice spacing $a \to 0$ in physical units),
\be
   \lim_{a \to 0} M_\mu(x) =  1 + ia A_\mu + O(a^2) \;,
\label{limit}
\eq
where $A_\mu$ is the usual hermitian gauge potential.
Thus, $M$ must tend asymptotically to the
unitary Wilson link variable \cite{wilson}, and in the
isotropic continuum limit $S$ will
be proportional to  $\Tr\left\{ F_{\mu \nu}F^{\mu \nu}\right\}$.  
For future reference, it is convenient to introduce lattice versions
of the covariant derivative and field strength tensor \cite{mack};
\be
        \D_\mu M_\nu(x) \equiv M_\mu(x) M_\nu(x+a \widehat{\mu})
                            - M_\nu(x) M_\mu(x+ a \widehat{\nu}) \; .
\label{field}
\eq
This natural definition reduces to $F_{\mu \nu}$ in the continuum.

More generally the
coefficients of $S$ are to be chosen to obtain
scaling, {\em id est} Green's function observables are to be independent
of the correlation length(s). 
We will demonstrate existence\footnote{Uniqueness would 
follow from the assumption that the continuum limit
of QCD is unique, since the light-front Hamiltonian approach requires
us to  take the continuum limit (\ref{limit}) in
two directions and then demand approximate Lorentz invariance.}
of an approximate scaling
trajectory, in the light-front Hamiltonian limit, by demanding
Lorentz covariance of boundstate solutions to the theory with $S$ 
truncated to a finite number of terms.
This empirical approach is similar in spirit to the Monte Carlo Renormalisation
Group \cite{wilson2}.
However, we find it more convenient to directly minimise violations of Lorentz
covariance in observables, as a function of couplings in $S$, rather than
try to perform explicit `block-spin' transformations.

\begin{figure}
\centering
\BoxedEPSF{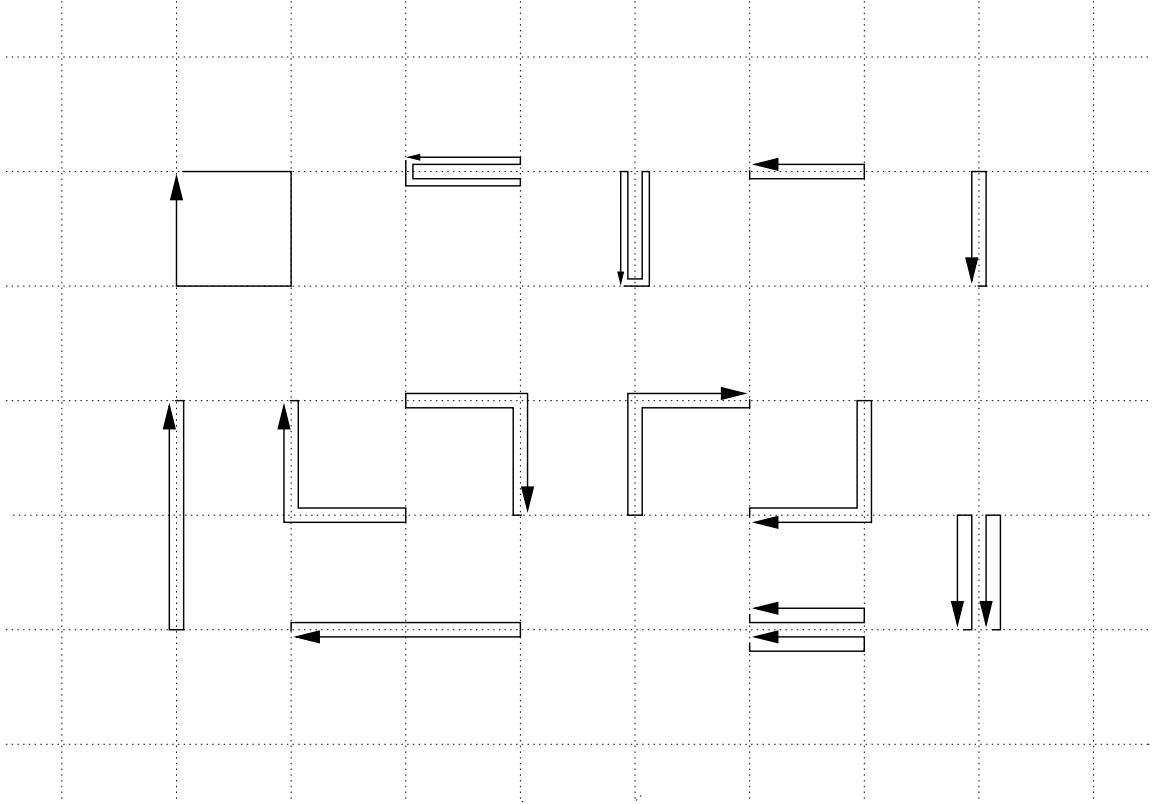 scaled 600}
\caption{All Wilson loops and their local products of total length
$\leq 4$. To those shown above one must add the different orientations
of each loop in $D$ spacetime dimensions and reversal of arrows 
(only the plaquette
gives a distinct loop in the latter case.) In the case of products of
Wilson loops, we have only shown the case where the loops lie exactly
on top of one another (local product).
\label{fig1}}
\end{figure}

The Wilson loops up to length 4 are illustrated in Fig.~\ref{fig1}. 
The quadratic loop 
$\Tr\left\{M_\mu(x) M_\mu^{\da}(x) \right\}$, 
{\em id est} the `mass squared' term of the $M_\mu(x)$ fields,
 evidently plays a special role. It must 
obviously have negative (tachyonic) 
coefficient in the continuum limit (\ref{limit}). 
If it has a positive coefficient, a `strong coupling' expansion of
the path integral (\ref{smear}) about 
$M_\mu(x)=0$ may be performed \cite{wein}. 
If the scaling trajectory passes through such a
region, Mack showed that a colour-dielectric picture of confinement
results \cite{mack,kogsus}. Accordingly, we shall refer to this as the {\em 
colour-dielectric regime}. There is   
evidence for $SU(2)$ that a partition
function of the form (\ref{smear}), subject to `block-spin' transformations,
does have renormalisation group trajectories  which pass into
the colour-dielectric regime at short enough correlation length \cite{pirner}. 
Therefore, the picture one should keep in mind is the following. 
If one decomposes $M_\mu(x)=HU$ into a hermitian matrix $H$ and a unitary 
matrix
$U$, in the continuum limit $a \to 0$,
$U$ is identified with Wilson's link variable.
$H = H_{0} + \tilde{H}$ gets a VEV $H_{0}$, while the fluctuation
$\tilde{H}$ becomes very heavy and decouples. Near the continuum limit, $H_0$
appears in the equations of motion like a generalised dielectric
constant \cite{mack}. In this regime the field $M_\mu(x)$ is tachyonic.
As the lattice spacing $a$ is
increased, a scaling trajectory in $S$ pushes one into a region of
positive mass squared 
for $M_\mu(x)$, where $H_0$ vanishes and $\tilde{H}$ is fully 
dynamical. The mass of $M_\mu(x)$ then increases with $a$. We will find further
evidence for this in our light-front Hamiltonian calculations later, where
it is rather 
crucial to work in a regime of positive mass squared for $M_\mu(x)$.

\subsection{Light-Front Continuum Limit.}

It is difficult to perform quantitative calculations with the
the dielectric lattice gauge theory described above because one must
search the multi-dimensional coupling constant space of $S$ for a
scaling trajectory. We will show that these problems
can be overcome in the light-front Hamiltonian limit.
To employ a light-front Hamiltonian formulation, it is desirable to
take the continuum limit in two directions.
Thus, in $D$ spacetime dimensions 
we will take the lattice spacing to zero in the $x^0$ and
$x^{D-1}$ directions while keeping the lattice spacing $a$ in
the $D-2$ remaining ``transverse'' directions constant.
Throughout, boldface will denote vectors in the transverse directions, 
${\bf b}$, with components $b^r$ where $r,s\in \left\{1,\ldots,D-2\right\}$.
Continuum Lorentz indices are denoted by $\alpha,\beta \in \{0,D-1\}$;
thus we write the inner product of two Lorentz vectors as 
$a^\mu b_\mu = a^\alpha b_\alpha -{\bf a\cdot b}$.
Transverse coordinates
${\bf x}$ and ${\bf y}$ will hereafter be restricted to
values corresponding to transverse lattice sites.
The covariant derivative (\ref{field}) in direction $x^\alpha$  becomes
\be
        \D_{\alpha} M_r({\bf x})  
        =  \left(\partial_{\alpha} +i A_{\alpha} ({\bf x})\right)
        M_r({\bf x})  
        -  i M_r({\bf x})   A_{\alpha}({{\bf x}+a\hat{r}}) \;,
\label{covdiv}
\eq
in the partial continuum limit.
One finds a transverse lattice action
\be
S  = \int dx^0 dx^3 \sum_{{\bf x}}  \left(
  \Tr\left\{ \D_{\alpha} M_r({\bf x}) \left(\D^{\alpha} 
M_r({\bf x})\right)^{\da}
\right\} - 
{1 \over 2G^2} \Tr\left\{F_{\alpha\beta}F^{\alpha\beta}\right\}  
- V_{{\bf x}}[M]\right) \;,
\label{lag}
\eq
where we have chosen
to normalise $M_r({\bf x})$ to have a canonical kinetic term, and repeated
indices are summed unless states otherwise. 
The dimensionful coupling $G^2(a)$ is such that 
$a^{D-2} G^2 \to g^{2}_{D}$ in the classical continuum limit $a \to 0$, 
where $g^{2}_{D}$ is the continuum gauge coupling in $D$ dimensions.
If we keep  only Wilson loops and their  products of 
total length 2 and 4 (which we justify in the next section), the 
effective potential 
$V_{{\bf x}}$ generically consists of 
Wilson loops among those shown in
Fig.~\ref{fig1}, but now understood to be in the 
remaining $D-2$ dimensional
transverse lattice and pinned at one point to ${\bf x}$.
%

The basis (\ref{lag}) for an effective QCD light-front
Hamiltonian was earlier proposed in Ref.~\cite{bard1}, 
motivated by considerations of two-dimensional
sigma models. 
Subsequently, an exploratory glueball calculation was performed \cite{bard2}
for some simple choices of $V$. Those authors speculated that one
might be able to find a scaling trajectory in $V$, corresponding to
QCD. We have extended the methods of Ref.~\cite{bard2}
in order to answer this question, and perform in this paper
quantitative first principles calculations. This allows us to confirm the
existence of an approximate scaling trajectory at large transverse
lattice spacing (at least in $2+1$ dimensions), 
and show that it yields the same
results for physical observables that are available from conventional
methods of studying QCD.

As mentioned above, to take best advantage of the 
simplifications on the light-front
afforded by linear link variables $M_r({\bf x})$,
it is quite important to
be able to quantise about $M_r({\bf x})=0$. 
Our calculations therefore must implicitly test a light-front Hamiltonian 
version of the colour-dielectric regime:

\vspace{3mm}
\noindent {\em In the light-front Hamiltonian limit of lattice action $S$, 
we search for a scaling trajectory 
by canonical 
quantisation about $M_r({\bf x})=0$, taken as the true groundstate.}
\vspace{3mm}

\noindent Once outside of the purported 
colour-dielectric regime,
when $a$ is sufficiently small and $M_r({\bf x})=0$ is no longer the minimum of
energy, we expect the zero mode structure of light-front quantisation to
become much more complicated \cite{prokh}. 


\section{Canonical Light-Front Quantisation.}
\label{clcq}

\subsection{High Energy Cutoffs}
\label{highecut}

We introduce light-front co-ordinates $x^{\pm} = x_{\mp} = 
(x^0 \pm x^{D-1})/\sqrt{2}$, for all two-vectors in the $(x^0,x^{D-1})$ plane,
and quantise by treating $x^+$ as canonical time. The variables
conjugate to $x^{\pm}$ and ${\bf x}$ are $k^{\mp}= (k^0 \mp k^{D-1})/\sqrt{2}$
and ${\bf k}$ respectively. Particles and 
antiparticles have $k^+ \geq 0$.
The transverse lattice theory will have divergences associated
with the $k^+ \to 0$ limit. This is a high $k^-$ light-front energy limit, 
as follows  from the dispersion relation
\be
k^- = {\mu^2 + |{\bf k}|^2 \over 2k^+} \label{disp}
\eq
for free particles of mass $\mu$. At finite transverse lattice spacing
$a$, this is the only source of divergence, since ${\bf k}$ is 
bounded by the Brillouin zone $\pi/a$. In fact, the theory is just a set
of coupled $1+1$ dimensional continuum field theories, and there are
only the mild free-field divergences characteristic of
two-dimensional super-renormalisable theories. For pure gauge theory,  
finiteness is ensured by a suitable
normal-ordering prescription.

At the level of free fields, it follows that 
if $\mu^2 >0$ the quantum vacuum is free of particles.
Because light-front momentum $k^+ \geq 0$ and is conserved,
the zero momentum vacuum state with even one massive ($k^+ = 0$) 
particle present is infinitely higher in energy than
one with no particles. 
In the interacting theory, the particles  must not condense either. In 
other words, one assumes that one is already expanding about the true
vacuum state, and all condensation effects are parameterised by
effective interactions in the Hamiltonian.
This scenario is precisely realised in the colour-dielectric 
regime; the dielectric fields $M_r({\bf x})$ are massive and 
$M_r({\bf x})=0$ is the
true vacuum. The non-trivial problem is to demonstrate a coupling
trajectory in this regime which restores all fundamental 
(unbroken) symmetries of the theory
that have been broken by the regulators.

For fixed total momentum $P^+$,
a large number of particles must carry a very large light-front energy, since
each particle is forced to carry a small positive $k^+$.
Although this would also be true of massive particles in
an equal-time quantisation, the effect
is much more pronounced in the light-front formulation, because of the peculiar
inverse relation between momentum and energy (\ref{disp}).
In particular, one notes that in the effective potential $V$,
higher powers of the field $M_r({\bf x})$ would couple precisely to those
components of the wavefunction with a large number of $M$-quanta. 
This is our  physical justification for truncation of $V$ to the shortest
Wilson loops. Higher loops may be added as greater accuracy is required
in the low-energy theory.

There are potentially massless degrees of freedom associated with the
continuum fields $A_{\pm} = (A^0 \pm A^{D-1})/\sqrt{2}$. The behaviour of these
modes is subject to the gauge fixing and regulation of the $k^+ = 0$ region, 
and is discussed in the next section.
We will impose periodic boundary
conditions in $x^-$ \cite{old}, with period ${\cal L}$, taking the longitudinal
momentum space continuum limit ${\cal L} \to \infty$ for fixed $a$. 
This gauge-invariant discretisation
is also a useful starting
point for the numerical investigation of the light-front Hamiltonian 
\cite{dlcq} (DLCQ).

\subsection{Gauge Symmetry}

The theory (\ref{lag}) possesses the fields 
$M_r(x^+, x^-,{\bf x})$, $A_{+}(x^+,x^-,{\bf x})$, and  
$A_{-}(x^+,x^-,{\bf x})$, together
%
%
with the unitary gauge symmetry at each transverse lattice site ${\bf x}$:
%
\begin{eqnarray}
        A_{\alpha}({\bf x}) & \to & U({\bf x}) A_{\alpha}({\bf x}) 
        U^{\da}({\bf x}) + {\rm i} \left(\partial_{\alpha} U({\bf x})\right) 
        U^{\da}({\bf x}) \\
        M_r({\bf x}) &  \to & U^{\da}({\bf x}) M_r({\bf x})  
        U({\bf x} + a\hat{r})   \; .
\end{eqnarray}
The gauge fixing of this theory may be performed following similar
treatments in Refs.~\cite{fix}. 
With periodic boundary conditions on $x^-$, we can pick the gauge
$\partial_{-} A_{-} = 0$.
This does not generate a Fadeev-Popov Jacobian 
since it is (almost) an axial gauge.
This leaves the zero modes
\be
{A}_{-}^{0}(x^+,{\bf x}) \equiv \int_{0}^{{\cal L}} 
dx^- {A}_{-}(x^+,x^-,{\bf x}) \; ,
\eq
which may be further gauge fixed to diagonal form 
\be
        {A}^{0}_{-ii}  \; , \;\;\;\; i \in \{1,2,...,N\} \; , 
        \;\;\;\; \sum_{i} {A}^{0}_{-ii}=0 \; , \label{diag}
\eq
by  $x^-$-independent gauge transformations. 
It is convenient to define
\be
        c_i = { {\cal L} \over 2 \pi} {A}^{0}_{-ii} \;  .
\eq
The gauge fixing to diagonal form (\ref{diag})
generates a Jacobian, which is just the invariant group measure for $SU(N)$,
proportional to
\be
        \prod_{i,j}\left( \sin{(c_i-c_j)\pi} \right)^2 \label{jac}
\eq
at each site ${\bf x}$ on the transverse lattice.
This produces a repulsion between the eigenvalues $c_i$. 
$A_{+}$ will turn out to be  a constrained 
field which may  be eliminated by its equation
of motion
in terms of the $M_r({\bf x})$'s and $c_i$'s. 

There are residual global gauge transformations
\be
        U_{ij} = \delta_{ij} {\rm e}^{{\rm i} \pi m_i x^- /{\cal L}} \;
        \; , \;\;\;\;  \sum_{i} m_i = 0 \; ,
\eq
where $m_i \in \{ \pm 2 , \pm 4, ....\}$; $c_i$ transforms as
\be
        c_i \to c_i - m_i/2 \;\;\;\; \mbox{(Gribov).} \label{grib}
\eq
This generates Gribov copies. 
For pure gauge theory there are also the center global
gauge transformations for each space direction $x^-$, ${\bf x}$.
For $x^-$ they are of the form
\begin{eqnarray}
        U_{ij} & = & \delta_{ij} {\rm e}^{{\rm i} 2 \pi x^- n / {\cal L}N} 
                \; , \;\;\;\; i,j<N \; , \\
U_{NN} & = &  {\rm e}^{-{\rm i} 2 \pi x^- (N-1) / {\cal L}N}  \; ,
\end{eqnarray}
for $n \in \{1, 2, \cdots , N-1\}$.
For $N=\infty$ we can ignore the last matrix element and think of
$\exp(2 \pi x^- s/{\cal L}) \cdot {\bf 1}$ for $0<s<1$. (More generally we will
ignore the extra $U(1)$ difference between $SU(\infty)$ and $U(\infty)$
when it does not matter.) $c_i$ transforms  as
\be
c_i \to c_i - s \;\;\;\;\; \mbox{(Centre).} \label{cent}
\eq
In the case that $x^1$, say, is compactified with  period $ L_{1}$, 
the center transformations are 
accordingly $\exp(2 \pi{\rm i} x^1 s/L_{1}) 
\cdot {\bf 1}$ for $0<s<1$. Under the latter, $M_r({\bf x})$ 
for 
$\hat{r} = (1,0)$ transforms as
\be
M_r({\bf x}) \to 
{\rm e}^{2 \pi {\rm i} s a /  L_{1}} M_r({\bf x}) \; .
\eq
Thus a Polyakov loop (winding mode) $W_{(p,0)}$, of winding number $p$ around
$L_{1}$, transforms as
\be
        W_{(p,0)} \to {\rm e}^{2 \pi {\rm i} p s}W_{(p,0)} \; , \label{center}
\eq
independent of $L_1$.

\subsection{Confinement and Zero Modes}

The final gauge-fixed action is
\be
S = S_{1} + S_{2} + S_{3} \;,
\eq
where
\begin{eqnarray}
        S_1  & = & \int dx^+ dx^- \sum_{{\bf x}}  
        {\rm Re}\left[ \Tr\left\{ \partial_{+} M_r({\bf x}) 
        \D_{-} M_r^{\da}({\bf x})
        \right\}\right] - V_{\bf x}[M]   \label{a1}  \\
        S_2 &= &   \int dx^+ dx^- \sum_{\bf x}  {1 \over G^2} 
        \Tr\left\{ \left(D_{-} A_{+}({\bf x})\right)^2 \right\} + \Tr \ 
        \left\{ A_{+}({\bf x})
        J^{+}({\bf x}) \right\}  \label{a2} \\
        S_3 &=&  \int dx^+ 
        \sum_{i,{\bf x}} {4 \pi^2 \over G^2 {\cal L}} (\partial_{+} 
        c_i({\bf x}))^2 \label{a3} \\
   J^{+}({\bf x}) & = & i \left[
        M_r({\bf x}) 
        \left(\D_- M_r({\bf x}) \right)^{\da}  - 
           \left(\D_- M_r({\bf x}) \right) M_r^{\da}({\bf x})+
        \right.  \nonumber \\ && \hspace{0.25in} \left.
           M_r^{\da}({\bf x} - a\hat{r}) \D_- M_r({\bf x} - a\hat{r})  - 
          \left(\D_- M_r({\bf x} - a\hat{r})\right)^{\da}
          M_r({\bf x} - a\hat{r}) \right] 
                \; . \label{jai}
\end{eqnarray}
Here $D_{-}\phi = \partial_{-}\phi + {\rm i}[A_{-},\phi]$.
The $A_+$ constraint equation of motion obtained from $S_2$ 
is then
\be
{2 \over G^2} D_{-}^{2} A_{+}  =  J^{+} \label{con} \;.
\eq
Formally inverting (\ref{con}) one derives the longitudinal
`Coulomb' interaction at site ${\bf x}$,
\be
        S_2   =  \int dx^+ dx^- \sum_{{\bf x}}  
                {G^2 \over 4} \Tr\left\{J^{+}({\bf x}) 
                (D_{-})^{-2} J^{+}({\bf x}) \right\}  \ .
\label{cou}
\eq
We expect this to produce linear confinement in the longitudinal direction,
because of the  classical nature of this effect for the
two-dimensional gauge theory at each lattice site.

The degrees of freedom $c_i$ are essentially winding modes around ${\cal L}$, 
since
\be
\Tr\left\{{\rm P} \, 
\exp\left( i \int_{0}^{{\cal L}} dx^- A_{-} \right)\right\}
= \sum_{i} {\rm e}^{2 \pi {\rm i} c_i}
\eq
in the present gauge. Therefore  the  classical linear confinement
in two dimensions should push these modes to infinite energy in the
${\cal L} \to \infty$ limit. 
We can see this more formally in the large-$N$ limit. 
The Schr{\"o}dinger representation of the kinetic term $S_3$ 
(\ref{a3}), 
\be
-{G^2 {\cal L} \over 16 \pi^2} {d^2 \over dc_{i}^{2}} \label{kin} \;,
\eq
evidently demonstrates that 
the modes $c_i$ 
become classical at $N=\infty$, since $G^2 N$ is held finite.
Introducing
\be
        J_{ij}^{+}(x^-,{\bf x}) = 
                \sum_{n=-\infty}^{\infty} \tilde{J}^{+}_{ij}(n,{\bf x}) 
                {\rm e}^{-2 \pi {\rm i} nx^- /{\cal L}} \;,
\eq
the longitudinal  interaction (\ref{cou}) becomes\footnote{The apparent 
divergence at $n=0$, $c_{i}=c_{j}$, can be
avoided by using residual gauge freedom
to gauge away ${A}^{0}_{+ii}(x^+)$ for each $i$ 
at a specific light-front time $x^+$, taken as the initial value 
surface $x^+ =0$ on which the constraint
equations are solved \cite{fix}.}
\be
G^2 {\cal L}  \sum_{n=-\infty}^{+\infty} 
{\tilde{J}^{+}_{ij}(n,{\bf x})\tilde{J}^{+}_{ij}(-n,{\bf x})
\over (n-c_i+c_j)^2} \;. \label{coul} 
\eq
The residual global gauge invariances, (\ref{grib}) and (\ref{cent}), 
mean that 
we can look for the classical minimum configuration in the fundamental domain
$0 < c_i <1$. From (\ref{coul}), we see that if 
$ \tilde{J}^{+}_{ij}(0,{\bf x}) \neq 0$,
this term contributes an energy of
order ${\cal L}$.
On the other hand, if $\tilde{J}^{+}_{ij}(0,{\bf x})$ vanishes, say
$\tilde{J}^{+}_{ij}(n,{\bf x}) \sim (n/{\cal L})^s$, 
a finite energy is only obtained
in the ${\cal L} \to \infty$ limit if 
$s>1/2$. In this case, only $n \sim O({\cal L})$ 
contributes to finite energy configurations and the $c_i$ are negligibly
small by comparison in the covariant
derivative $D_{-}$. (States on which
$\tilde{J}^{+}_{ij}(0,{\bf x})$ vanishes  also produce the necessary factor
of $N$ to make matrix elements of (\ref{coul}) finite in the large $N$
limit; see Section~\ref{focsec}.)  This argument treats the currents $J^+$ as
essentially classical sources, and could be upset if quantum
fluctuations of $M_r({\bf x})$ led to complete
screening of the linear potential (\ref{cou}). 
Provided no condensation of the field $M_r({\bf x})$ takes place, as is 
assumed in the colour-dieletric regime, this cannot occur.


The above arguments apparently 
justify use of the naive light-front gauge $A_{-}=0$
in the original work \cite{bard1,bard2}. One possible caveat concerns
the $O({\cal L})$ contributions of $c_i$ to the light-front energy, 
from the longitudinal Coulomb interactions (\ref{coul}) between link variables.
Although these are infinite as ${\cal L} \to \infty$, as demonstrated above,
a principle value prescription operates between
Coulomb self-energy and  Coulomb exchange at zero $k^+$ momentum transfer,
such that linear divergences cancel. It is plausible that the $c_i$, 
being relevant near the zero momentum
region, might affect this cancellation in such a way as to alter the residual
finite part. Very recently, some numerical evidence for this has been given for
scalar $SU(2)$ matter \cite{pauli}
(in the case
of $SU(2)$ there is only one value of $i$). 
In Appendix~\ref{appendixa} we give an argument which shows that the only 
effect of the
zero modes $c_i$ is to finitely renormalise the link-field mass $\mu^2$ via the
singular Coulomb interaction (\ref{coul}). This is of no consequence for our 
approach, since $\mu$ is a renormalised parameter to be determined by 
physical measurements anyway.
We will therefore proceed by dropping
the variables $c_i$ from now on.

The charges $\tilde{J}^+(0,{\bf x})$ 
generate $x^-$-independent gauge transformations
before gauge fixing, and the requirement that they 
vanish for finite energy states is the statement of colour confinement
for the transverse lattice formulation at fixed $a$ \cite{bard2}. The mechanism
for the confining linear potential in longitudinal and transverse directions is
somewhat different in each case however. For the longitudinal direction
it arises as the Coulomb potential (\ref{cou})
for the $1+1$-dimensional gauge field theory at each site ${\bf x}$.
In the transverse directions it arises through the Wilson-Kogut-Susskind
argument for lattice gauge theories \cite{wilson,kog}. 
For colour singlet at each
${\bf x}$, we must build a colour string with the (massive) $M_r({\bf x})$ link
fields, the energy being proportional to the number of links in the string.
Note however that the colour singlet requirement came from the continuum
longitudinal dynamics.
In later sections, we will
show how one can verify these intuitive arguments by explicit calculation
of the transverse and longitudinal string tensions.

\subsection{Energy and Momentum}

With the considerations of the last subsection in place, only
$M_r({\bf x})$ remains as a physical degree of freedom.
We  derive the total light-front momentum and energy 
 $P^\mu = \int dx^- \sum_{{\bf x}} \theta^{+ \mu}$,
\begin{eqnarray}
        P^+ & = & 2 \int dx^- \sum_{{\bf x}, r} \Tr  
                        \left\{ \partial_- M_r({\bf x})  
                \partial_- M_r({\bf x})^{\da} \right\} \label{mom} \\
        P^-  & = & \int dx^- \sum_{{\bf x}}  
                V_{{\bf x}}[M] - {G^2 \over 4} \left(\Tr\left\{ 
              J^{+}({\bf x}) \frac{1}{\partial_{-}^{2}} J^{+}({\bf x}) \right\}
            -{1 \over N} 
        \Tr\left\{ J^+ ({\bf x}) \right\} {1 \over \partial_{-}^{2} }
     \Tr\left\{ J^+({\bf x}) \right\} \right)
        \; .  \label{energy} 
\end{eqnarray}
with $J^{+}({\bf x})$ now given by (\ref{jai}) with $\D_{-} \to \partial_{-}$.
The $1/N$ correction to the Coulomb term in (\ref{energy})
is necessary at large-$N$ for $SU(N)$ to 
ensure that the longitudinal 
gluon $A_+$, having no colour singlet part, does not couple to 
gauge singlet states. It will turn out to be important only for two-link
states at non-zero transverse momentum ${\bf P}$.

Given the absence of modes $c_i$, we 
recover the residual local gauge symmetry generated by 
$\tilde{J}^{+}(0,{\bf x})$
\be
        M_r({\bf x}) \to  U^{\dagger}({\bf x}) M_r({\bf x})  
                U({\bf x} + a\hat{r})  \label{red}
\eq
with $U({\bf x})$ independent of $x^-$.
Following the discussion of the previous subsection, 
the light-front energy (\ref{energy}) is
not finite unless the associated charge vanishes at  each site ${\bf x}$: 
$\tilde{J}^{+}(0,{\bf x}) = 0$.
Thus, only states invariant under the residual gauge symmetry
(\ref{red}) will have finite energy. 

\subsection{Fock Space}
\label{focsec} 

We impose canonical commutation relations at $x^+ = 0$
\be
        \left[M_{r,ij}(x^-,{\bf x}), 
        \left(\partial_- M_{s,kl}(y^-,{\bf y})\right)^\da\right]
        = {1 \over 2} \delta_{il}\,\delta_{jk}\, \delta (x^- -y^-)
        \,{\bf \delta_{x, y}} \,\delta_{r,s} \;.
\eq
We work in longitudinal momentum space and transverse position space
\begin{eqnarray}
 M_r(x^+=0,x^-,{\bf x})   &=&  
        \frac{1}{\sqrt{4 \pi }} \int_{0}^{\infty} {dk \over {\sqrt k}}
        \left( a_{-r}(k,{\bf x})\, e^{ -i k x^-}  +  
        a^{\da}_r(k,{\bf x})\, e^{ i k x^-} \right)   \; ,\\
   \left[a_{\lambda,ij}(k,{\bf x}), 
        \left(a_{\rho,kl}(\tilde{k}, {\bf y})\right)^{\da}\right] 
        & = & \delta_{ik}\, \delta_{jl}\, \delta_{\lambda \rho}\, 
        {\bf \delta_{x, y}}\,\delta(k-\tilde{k}) \;, \\
   \left[a_{\lambda,ij}(k,{\bf x}),
        a_{\rho,kl}(\tilde{k}, {\bf y})\right] & = & 0 \;.
\end{eqnarray}
Here, $\left(a_{\lambda,ij}\right)^{\da} = (a^{\da}_{\lambda})_{ji}$ 
and $\lambda, \rho \in \{ \pm 1, \ldots \pm (D-2)\}$
For clarity, we omit the $+$ superscript on light-front momentum $k^+$ 
hereafter. 
The Fock vacuum $\left|0\right\rangle$ satisfies 
$\mbox{:}P^-\mbox{:}\left|0\right\rangle = 
\mbox{:}P^+\mbox{:} \left|0\right\rangle =0$ 
and  is the physical vacuum.
The Fock space operator $a_{\lambda}^{\da}(k,{\bf x})$ 
creates a mode with longitudinal momentum
$k$ on the link emanating from site ${\bf x}$ in transverse direction
$\widehat{\lambda}$.
This Fock space is already diagonal in $P^+$, and serves as a basis
for finding the energy levels of $P^-$. Since the glueball boundstates
do not interact with one another at $N = \infty$, this 
in turn will diagonalise
the invariant mass operator ${\cal M}^2 = 2 P^+ P^- - {\bf P}^2$ in the
frame ${\bf P} = 0$. 

\begin{figure}
\centering
\BoxedEPSF{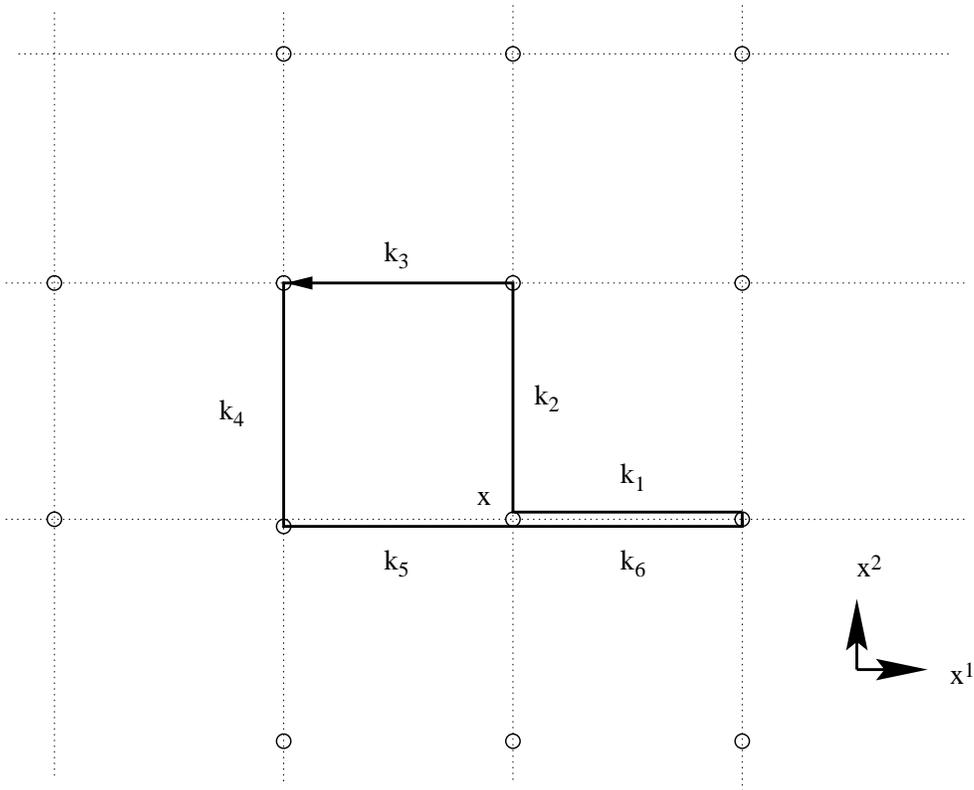 scaled 600}
\caption{ An example of a length 6 loop on the transverse lattice, showing
also the longitudinal momentum carried by  each link,
corresponding to the Fock state (\ref{typ}).
\label{fig2}}
\end{figure}

The combinations singlet under
(\ref{red}) are the `string' states which form closed flux loops on the
transverse lattice, {\em exempli gratia}
\begin{eqnarray}
        && \Tr\left\{ a_{-1}^{\da}(k_1,{\bf x}+a\widehat{1})\, 
        a_2^{\da}(k_2,{\bf x})\, a_{-1}^{\da}(k_3,{\bf x}+a \widehat{2}) 
        \right. \nonumber \\
        && \hspace{0.25in} \cdot \;\left.  
        a_{-2}^{\da}(k_4,{\bf x}+a \widehat{2}-a\widehat{1}) \,
        a_{1}^{\da}(k_5,{\bf x}-a \widehat{1})\,
        a_{1}^{\da}(k_6,{\bf x})\right\} \,  |0\rangle 
         \label{typ}
\end{eqnarray}
as illustrated in Fig.~\ref{fig2}.
The longitudinal momenta $k_m$ are
unconstrained except for $k_m > 0$ and conservation,
 $\sum_{m=1}^{6} k_m = P^+$ in the above example.


\section{Structure Functions.}
\label{structure}

A standard gauge invariant 
definition in the continuum of the gluon distribution function $G(x)$
in  a hadron $|\Psi(P^+)\rangle$ is \cite{soper}
\be
G(x) = {1 \over 2\pi x P^+} \int dy^- {\rm e}^{-{\rm i} xP^+ y^-}
\langle\Psi(P^+)|\,
F_{a}^{+r}(0,y^-,{\bf 0})\, {\cal O}_{ab}\, F_{b}^{+r}(0,0,
{\bf 0})\, |\Psi(P^+)\rangle \label{glue}
\eq
where $F_{a}^{\mu \nu}(x^+,x^-,{\bf x})$ is the field strength
tensor with colour index $a,b,c \in \{1, \ldots, N^2 -1\}$ and
\be
{\cal O} = {\cal P}\exp\left( {\rm i} \int_{0}^{y^-} dz^-
A_{c}^{+}(0,z^-,{\bf 0})\, t_{c}\right) \;. 
\eq
Here, the $t_c$ are the $SU(N)$ generators in the adjoint representation.
In the light-front gauge $A_{-}=0$ this definition (\ref{glue})
reduces to the expectation value of the gluon
number operator. Note that the answer is in fact independent of the
choice of ${\bf x}$. 
Since the colour-dielectric 
formulation deals with a set of collective variables
rather than  with gluons, the transformation from one to the
other being unknown explicitly, we cannot directly measure the gluon 
distribution.
We recall however from (\ref{field}) and (\ref{covdiv}) that 
$\D_{-}M_r({\bf x})$
reduces to the field strength $F_{- r}$ in the continuum limit. 
It seems sensible therefore to make the substitution $F_{- r} \to
\D_{-}M_r({\bf x})$ in the definition
(\ref{glue}). We call the resulting gauge invariant
distribution $G_{d}$ to avoid confusion. In the
continuum limit $G_{d}$ reduces to $G(x)$ but in general they are different.
Since $M_r({\bf x})$ are lattice variables,
$G_{d}$ will evolve with the renormalisation scale defined by the edge
of the Brillouin zone $\Lambda_\perp = \pi /a$.
In the light-front gauge $A_{-} = 0$ we have 
\begin{eqnarray}
G_{d}(x,\Lambda_{\perp}) & = &
{1 \over 2\pi x P^+} \int dy^- {\rm e}^{-{\rm i} x P^+ y^-}
\langle\Psi(P^+)|
        \Tr\left\{ \partial_{-} M_r \partial_{-} M^{\dagger}_r \right\}
|\Psi(P^+)\rangle \\
& = &
\langle\Psi(P^+)|\, {a}^{\dagger}_{\lambda, ij} (x P^+,{\bf x})\,
{a}_{\lambda, ij}(xP^+,{\bf x})\, |\Psi(P^+) \rangle  \label{str}
\end{eqnarray}
Thus $G_{d}$ has a simple interpretation:
the expectation value of the number operator for link fields of
longitudinal momentum fraction $x$. 
It trivially satisfies the momentum sum rule
\be
\int_{0}^{1} dx \ x G_d(x) = 1.
\eq
and the modes of $M_r({\bf x})$ may be interpreted as `fat' partons with 
finite transverse extent.
Bjorken scaling violations due to this finite extent will manifest themselves
through the evolution with $\Lambda_\perp$ along the scaling trajectory.
{}From the discussion of Section~\ref{euclid},
 one expects the scaling
trajectory to pass from large to small values of parton mass
$\mu^2$ as the lattice spacing $a$ is reduced. Since $a$ is a
measure of the transverse size of the parton, the parton becomes
fatter as it becomes heavier.
As $\Lambda_\perp$ is increased ($a$ decreased) the
glue distribution $G_{d}$ is expected to evolve to smaller $x$. 
This follows simply from
the fact that the small $x$ region
becomes enhanced the  lighter the  parton masses, since the free light-front
kinetic energy (\ref{disp}) is  $ \sim \mu^2 / x$ for a parton of
momentum fraction $x$. We explicitly calculate structure functions
later.

\section{The Large-$N$ Limit}
\label{largen}

In pure gauge theory,
the corrections to the large-$N$ limit usually form an asymptotic series in
$1/N^2$. For our problem, a dramatic simplification of the relevant Fock space
takes place when one
neglects these corrections.
Firstly, in Ref.~\cite{dalley1} and \firstpaper{} it
was shown that in the colour-dielectric regime,
the Hilbert space of ${\bf P} =0$
Wilson loops on the transverse lattice,
together with the  light-front Hamiltonian which acts upon them, 
undergo dimensional reduction; they
may be
replaced by an equivalent problem where all links in a given transverse
direction are identified. One effectively works on a transverse
lattice with one link in each direction and periodic boundary 
conditions. This is a manifestation of Eguchi-Kawai
reduction \cite{kawai} in the light-front Hamiltonian framework. 
One may drop the argument ${\bf x}$ in all previous expressions,
\be
        M_r(x^-,{\bf x}) \to M_r(x^-) \;\ , \;\
        a_{\lambda}(k, {\bf x}) \to a_{\lambda}(k) \;. \label{reduced}
\eq
The proof of the dimensional 
reduction (\ref{reduced}) does not encounter any phenomena
analogous to center $U(1)$ breaking \cite{break}; 
however one must recall that the
colour-dielectric regime, on which the proof is based, is only
expected at large transverse lattice spacing.
In this way, in $D$ spacetime dimensions one ends up
having to solve two-dimensional large-$N$ gauge theories coupled to $2^{D-2}$ 
complex adjoint scalars. The light-front quantisation of this kind of
two-dimensional gauge theory has recently been developed in Refs.~\cite{model},
whose mathematical analysis we follow closely.\footnote{The 
dimensional reduction
discussed in those papers, which is an approximation,
is not to be confused with the large-$N$ 
reduction discussed here, 
which is exact.} 
 
The Hilbert space is
further simplified by the fact that the light-front Hamiltonian propagates
Wilson loops without splitting or joining at 
large-$N$ \cite{thorn}; the amplitude for this is order $1/N$. Thus the
light-front Hamiltonian may be  studied in the space of connected Wilson loops,
that is, states containing one colour trace, {\em exempli gratia} (\ref{typ}). 

The explicit light-front Hamiltonian (\ref{energy}) for the 
frame ${\bf P} =0$ now becomes equivalently $(D=4)$
\begin{eqnarray}
 P^-  & = & \int dx^- 
 - {G^2 \over 4} \Tr\left\{  J^{+} \frac{1}{\partial_{-}^{2}} J^{+} \right\} 
\nonumber \\
%
%
&& -{\beta \over Na^{D-2}} \Tr\left\{ M_{2}^{\da} M_{1}^{\da}
M_{2} M_{1} 
+ \mbox{hermitian conjugate} \right\} +  \sum_r  
 \mu^2  \Tr\left\{M_r M_r^{\da}\right\} \nonumber \\ 
&& + \sum_r  {\lambda_1 \over a^{D-2} N}
\Tr\left\{ M_r M_r^{\da}
M_r M_r^{\da} \right\} +
 {\lambda_2 \over a^{D-2} N}\sum_r  
\Tr\left\{ M_r M_r
M_r^{\da} M_r^{\da} \right\} \nonumber \\
&& + \sum_r   {\lambda_3 \over a^{D-2} N^2} 
\left( \Tr\left\{ M_r M_r^{\da} \right\} \right)^2
+  {\lambda_4 \over a^{D-2} N}  
\sum_{\sigma=\pm 2, \sigma^\prime = \pm 1}
        \Tr\left\{ 
M_\sigma^{\da} M_\sigma M_{\sigma^\prime}^{\da} M_{\sigma^\prime} \right\} 
        \nonumber\\
&& +  {4 \lambda_5 \over a^{D-2} N^2} 
\Tr\left\{ M_1 M_1^{\da} \right\}\Tr\left\{ M_2 M_2^{\da} \right\} \ ,
        \label{redham}\\ 
 J^{+} &=&  i \left(
M_r \stackrel{\leftrightarrow}{\partial}_{-} 
M_r^{\da}  + M_r^{\da} 
\stackrel{\leftrightarrow}{\partial}_{-} M_r   \right) \; .
\end{eqnarray}
%
In Section~\ref{lorentz}
we will show how ${\bf P} \neq 0$ may be treated in the
same dimensionally reduced framework, by adding appropriate
phase factors to matrix elements of (\ref{redham}) in the dimensionally
reduced
Fock basis. We have shown in (\ref{redham}) 
terms of the effective potential 
corresponding to Fig.~\ref{fig1} on the transverse lattice, with coefficients 
$\mu^2, \lambda_1, \lambda_2, \lambda_3, \lambda_4$, $\lambda_5$, and $\beta$. 
These parameters are to be fixed by a scaling analysis. 
Note that the multiple Trace operators coupling to
$\lambda_3$ and $\lambda_5$ {\em do}
contribute in leading order of $N$, but only on 2-parton Fock states.
The dimensional reduction in the large-$N$ limit
was not discussed by the authors of Refs.~\cite{bard1,bard2}, 
but was used implicitly in the calculations performed there. 

The fact, that the large-$N$ limit of the light-front Hamiltonian 
simplifies the calculations of physical observables  in this way is to
be contrasted with other quantisation schemes.
A  path integral evaluation of Green's functions
\be
	\left\langle G[A_\mu] \right\rangle = 
	      \int DA_\mu\, G[A_\mu]\, 
		\exp\!\left(- \int dx\, S[A_\mu]\right)
\eq
should be dominated exactly at $N= \infty$ by an $x$-independent
classical saddle point configuration of the 
fields $A_\mu$, the `master field' \cite{witten}. 
This is motivated by the factorisation property
\begin{eqnarray} 
\lefteqn{\left\langle \Tr \left\{A_{\mu_1}A_{\mu_2}\cdots 
A_{\mu_n}\right\} \Tr \left\{A_{\mu_{n+1}}A_{\mu_{n+2}}
\cdots A_{\mu_m}\right\}\right\rangle } \nonumber \\
&=&\left\langle{\rm Tr}\left\{ A_{\mu_1}A_{\mu_2}\cdots 
A_{\mu_n}\right\}\right\rangle\, \left\langle \Tr \left\{A_{\mu_{n+1}}
A_{\mu_{n+2}}
\cdots A_{\mu_m}\right\}\right\rangle \nonumber \\
&& + {1 \over  N^2} 
\left\langle\Tr \left\{A_{\mu_1}A_{\mu_2}\cdots 
A_{\mu_n}\right\}\Tr \left\{A_{\mu_{n+1}}A_{\mu_{n+2}}
\cdots A_{\mu_m}\right\}\right\rangle_c  \label{fac}
\end{eqnarray}
where $\left\langle\cdots\right\rangle_c$ 
is the connected amplitude. Eqn.~(\ref{fac}) 
can be understood by the
usual planar diagram counting rules \cite{hoof}, which attach a factor
$N^{2-2G-P}$ to a diagram with $G$ handles and $P$ boundaries. 
The expectation value of two
Wilson loops has a term of order $N^2$, from the disconnected disks ($P=1$
for each) covering
each loop, plus  a term of order $N^0$ from the connected amplitude, 
the cylinder connecting the two loops. 
Thus  at $N=\infty$, where Eguchi-Kawai reduction in the lattice
theory takes place, it appears that 
one can only calculate disconnected Green's functions and the vacuum energy.
In an equal-time Hamiltonian approach,  eigenvalues are also dominated by the
vacuum contribution of order $N^2$ (amplitude $N$ to create or
annihilate a zero momentum Wilson loop from the vacuum). 
It is therefore not possible to extract
the physical spectrum at $N= \infty$, since this involves contributions down 
by order $1/N^2$ compared to the vacuum contribution. Physical states 
interact with the vacuum by emitting or absorbing zero-momentum 
Wilson loops  with amplitude
of order $N^0$ (amplitude $1/N$ to split or join and amplitude $N$ to 
emit or absorb from the vacuum).

It is then clear why the transverse lattice 
light-front Hamiltonian framework is uniquely suited for 
calculations of observables---as opposed to vacuum 
structure---in the large $N$ limit. 
The triviality of the light-front vacuum
implies that there is no order $N^2$ contribution. Non-trivial
contributions begin at order $N^0$, corresponding to the free string-like
propagation
of a Wilson loop. In this way the spectrum of physical states is
directly computable at $N = \infty$, where a dimensional reduction in the
number of degrees of freedom takes place in the lattice theory. 
However, unlike the path integral vacuum problem where a reduction to zero
dimensions takes place, in the light-front transverse lattice approach
dimensional reduction takes place only in the transverse directions. 
The statement of large-$N$ (Eguchi-Kawai) reduction in $D$ spacetime dimensions
for physical 
observables, rather than vacuum structure, is
that {\em one can solve for observables in 
a field theory of $D-2$ fewer dimensions in the
large $N$ limit}.


\section{String Tensions}
\label{tensionsection}

\subsection{Transverse Tension}
\label{trans}

To measure the string tension $\sit$ in the ${\bf x}=(x^1,x^2)$ directions, 
consider a transverse lattice with ${\bf n}= (L_1 /a, L_2 /a)$
links and periodic boundary conditions. 
We construct a basis of Polyakov loop 
winding modes in Fock space in the frame ${\bf P} = 0$ 
that wind once around this lattice in both the $x^1$ and
$x^2$ directions. 
The shortest such loop has length $\left|{\bf n}\right|$ in lattice
units. One
may extract from the lowest invariant mass eigenvalue ${\cal M}^2$ 
the lattice string tension $a\,\sit$ via
\be
     {{\cal M}^2} \approx a^2 \sit^2 \left|{\bf n}\right|^2 \label{nnn}
\eq
for sufficiently large $\left|{\bf n}\right|$. 
The form (\ref{nnn}) will be imposed as one of
the scaling requirements, {\em id est} $\sit$ must not to vary with $a$ or
direction ${\bf n}/|{\bf n}|$.
Because of large-$N$ reduction, this procedure is equivalent to using
Polyakov loops $W_{{\bf n}}$ of winding number ${\bf n}$, 
on the transverse lattice with one link in each direction.
The total winding number in each direction is conserved due to center
symmetry (\ref{center}). In the large-$N$ limit, we study single Polyakov
loops for the same reason that splitting of Wilson loops $W_{(0,0)}$
is suppressed by $1/N$. Thus ${\bf n}$ labels disconnected
sectors of the Hilbert space.

In sectors of non-zero winding $\left|{\bf n}\right| \neq 0$,
there are further operators one could 
add to the effective potential $V[M]$ at the same level of approximation.
They will in general be necessary for finiteness, as well
as to achieve scaling in these sectors.
In particular, when the Hamiltonian (\ref{redham})
is put into normal-order, two kinds of quadratic terms are generated,
with divergent coefficients $c_m$ and $c_w$, in the
${\cal L} \to \infty$ limit:
\begin{eqnarray}
c_m \int_{0}^{\infty} {dk \over k} 
\Tr\left\{ a^{\da}_{\lambda}(k) a_{\lambda}(k) \right\} ; \label{massren}\\
{c_w \over N} 
\int_{0}^{\infty} {dk \over k} \Tr\left\{ a^{\da}_{\lambda}(k)
 \right\} \Tr\left\{ a_{\lambda}(k) \right\} \;.
\label{wmass}
\end{eqnarray}
The first (\ref{massren}) is a link-field mass self-energy, and is dealt
with by straightforward mass renormalisation, leaving 
the finite part $\mu^2$. The
second (\ref{wmass}) 
is suppressed by $1/N$ on all states except 
$\Tr\left\{ a^{\da}_{\lambda}(P^+) \right\}$,
the one-parton state of winding number $\widehat{\lambda}$.
To avoid the divergence of the mass of this state, one must add a
new counterterm to $V[M]$ given by
\be
-{1 \over N} 
 \Tr\left\{ M_r \right\} \Tr\left\{ M_r^{\da} \right\} \;,
\label{new}
\eq
with an appropriate coefficient to cancel $c_w$. It is then possible to 
have a finite part  left over, which will  
appear as a new parameter in the theory. In this case one loses 
one unit of predictive
power from the measurement of energy levels in  the winding number
$\pm 1$ sector.
Evidently, there are many similar multiple Trace
operators one can add to
the Hamiltonian, which act in this  way on ${\bf n} \neq 0$ 
sectors. They 
are quite analogous to the multiple trace
operators, of which $\lambda_3$ is an example (\ref{redham}), 
which can appear in the 
${\bf n}=0$ sector (in that case winding number is conserved `in each Trace').
However, (\ref{new}) is the only one required on the grounds of
finiteness when the Hamiltonian (\ref{redham}) is put into normal order.
In a winding ${\bf n}$ sector,
multiple Trace operators that contribute in leading order of $N$ must
contain at least $\sum_r |n^r|$ powers of $M_r({\bf x})$.
Since we try to extract $\sit$ from large $\left|{\bf n}\right|$, 
we will neglect this possible proliferation of parameters to a
first approximation, since we neglect higher powers
of $M_r({\bf x})$ in the effective potential $V[M]$.

\subsection{Heavy Quark Potential}
\label{long}

In addition to using Polyakov loops of nonzero winding number in the
transverse direction, we can measure
the string tension by from the heavy quark potential~\cite{burkardt}.  
The advantage
of this approach is that the heavy quark potential can be measured in
the $x^{D-1}$ as well as in the transverse directions.  The main disadvantage
is that this quantity is numerically difficult to compute.  

For convenience, 
we will start with a heavy scalar field $\phi(x^+,x^-,{\bf x})$ of
mass $\rho$ in the fundamental representation of the colour group
(one can just as well start with the Dirac equation and take the
heavy limit, the result is the same). 
In addition to the pure glue `link-link' interactions, 
the most general action to fourth
order in the fields is
\be
  \sum_{\bf x} \int dx^- \, dx^+\, 
        \left(D_\alpha \phi\right)^\da D^\alpha \phi 
        - \rho^2 \phi^\da \phi - \frac{\rho \tau}{a^{D-3} N}
        \phi^\da \left(M_{r}^{\da} M_{r}+M_{r} M_{r}^{\da}
        \right)\phi
\eq
%
%
where
\be
                D_\alpha \phi^\da = \partial_\alpha \phi^\da
-i \phi^\da A_\alpha \; .
\eq
The heavy field contributes to the gauge current $J^\alpha$ 
in Eqn.~(\ref{jai}) 
\be
        J^\alpha_{\rm heavy} = -i \left(D^\alpha \phi\right)\phi^\da
         +i \phi \left(D^\alpha \phi\right)^\da \; ;
\eq
note that $\phi \phi^\da$ is an $N\times N$ colour matrix.
Since we are interested in the 
$\rho\to \infty$ limit, we have not included any hopping terms
like $\phi^\da({\bf x}) M_r({\bf x})\phi({\bf x}+a \widehat{r})$
and $\phi({\bf x})$ is pinned to site ${\bf x}$ of the lattice.
Also, we have not included any ``pinching terms'' like 
$\phi^\da \phi\phi^\da \phi$ which, for large $N$, acts only
on states consisting of two $\phi$'s.

Now for some kinematics:  
Let $P^\alpha_{\rm full}$ represent the full 2-momentum
of a system containing $h$ heavy particles.  
It is convenient to split the full momentum into a ``heavy'' part
plus a ``residual'' part $P^\alpha$,
\be
        P^\alpha_{\rm full} = \rho h v^\alpha + P^\alpha \; , 
        \label{resid}
\eq
%
where $v^\alpha$ is the covariant velocity of the
heavy quarks, $v^\alpha v_\alpha =1$.  
The full invariant mass squared (${\bf P} = 0$) is
\be
 {\cal M}^2 = 2 P_{\rm full}^+ P_{\rm full}^- =
      \left(h\rho\right)^2 +2 h \rho v^+ P^- +
        2 P^+\left( P^- + h \rho v^- \right)
\eq
Our choice of $v^+$ is arbitrary and it is convenient to choose it
such that $P^+=0$.  Consequently, $v^+ P^-$ is just the shift of the 
full invariant mass $\cal M$ due to the interactions:
\be
  {\cal M} = h \rho +v^+ P^- +O\left(1/\rho\right)\; .
\eq
Thus, $v^+ P^-$ is the usual energy associated with
the heavy quark potential; this will be our Hamiltonian.

We define creation-annihilation operators associated
with the heavy field:
\be
  \phi = \frac{1}{\sqrt{4 \pi}}\int_{-\infty}^\infty 
        \frac{dk}{\sqrt{\rho v^+ + k}} \left(
         b(k)\,{\rm  e}^{-i v_\alpha x^\alpha \rho-i k x^-}
         + d^\da(k)\,{\rm  e}^{i v_\alpha x^\alpha \rho+i k x^-}\right)
           \label{phi} \; .
\eq
%
The creation/annihilation operators $b(k)$ and $d(k)$ have the 
usual commutation relations. 
The ${\rm  e}^{i \rho v_\alpha x^\alpha}$ term removes an 
overall momentum $\rho v^\alpha$ from the 2-momentum.
The contribution of the heavy particle interactions to the 
``residual'' momenta $P^\alpha$ in Eqns.~(\ref{mom}) and (\ref{energy}) 
are, to leading order in large $\rho$ and $N$,
\begin{eqnarray}
     P^+_{\rm heavy} &=& \int_{-\infty}^\infty dk \,k
                 \left(b^\da(k) b(k)+\mbox{:}d(k)d^\da(k)\mbox{:} 
\right) \label{pplus} \\
     P^-_{\rm heavy} &=& -\int_{-\infty}^\infty dk \, \frac{k}{2 {v^+}^2}
                 \left(b^\da(k) b(k)+\mbox{:}d(k)d^\da(k)\mbox{:} 
\right)\nonumber \\
         & & - \frac{G^2}{4 \pi} \int dk_1 \, dk_2 \, dk_3 \,dk_4
                \frac{\delta(k_1 +k_2-k_3-k_4)}{(k_1-k_3)^2} 
                 b^\da(k_1) d^\da(k_2) d(k_4) b(k_3) \nonumber \\
         & & + \frac{1}{8 \pi} \int \frac{dk_1 \, dk_2 \, dk_3 \,dk_4}
                {\sqrt{k_2 k_3}} \left[
        \delta(k_1 +k_2-k_3-k_4) {\cal H}_{2\to 2}
    \right. \nonumber\\ & & \hspace{0.5in}\left.+ 
        \delta(k_1 -k_2-k_3-k_4)\left( {\cal H}_{1\to 3}+ {\cal H}_{3\to 1}
               \right)\right]
\end{eqnarray}
with
\begin{eqnarray}
 {\cal H}_{2\to 2}&=& \left(
               -\frac{G^2\left(k_2+k_3\right)}{(k_2-k_3)^2} 
                +\frac{\tau}{a^{D-3}N}
       \right) \nonumber \\ & & \hspace{0in}\cdot\left(
             \mbox{:} d(k_1)a_\lambda(k_2) a^\da_\lambda(k_3) 
d^\da(k_4)\mbox{:}+
              b^\da(k_1)a^\da_\lambda(k_2) a_\lambda(k_3) b(k_4)
        \right)  \\
{\cal H}_{1\to 3}&=&  \left(
               \frac{G^2\left(k_2-k_3\right)}{(k_2+k_3)^2} 
                +\frac{\tau}{a^{D-3}N}
       \right) \nonumber \\ & & \hspace{0in}\cdot\left(
             \mbox{:} d(k_1)a^\da_\lambda(k_2) a^\da_{-\lambda}(k_3) 
d^\da(k_4)\mbox{:}+
              b^\da(k_4)a^\da_\lambda(k_3) a^\da_{-\lambda}(k_2) b(k_1)
        \right) \\
{\cal H}_{3\to 1}&=&{\cal H}_{1\to 3}^\da \; .
\end{eqnarray}
%

As an example, let us construct a state with two heavy sources with rest
frame separation $L$ in $x^{D-1}$ (this corresponds to an $x^-$ separation 
of $2 v^- L$):
\begin{eqnarray}
  |2\rangle &=& \frac{2 \rho v^+}{\sqrt{N}}\int dx^- \,
      \phi^\da \!\left(x^+,x^- + L v^-\right) \,
      \phi\!\left(x^+,x^- - L v^- \right) |0 \rangle 
      {\rm  e}^{-2 i \rho v_\alpha x^\alpha} \\
    &=& \frac{1}{\sqrt{N}}\int_{-\infty}^\infty dk\, b^\da(k) d^\da(-k)
       {\rm  e} ^{2 i L v^- k} \; .
\end{eqnarray}
The associated energy shift is
\be
  \frac{\langle 2|v^+ P^- |2\rangle}{\langle 2|2\rangle} =
      \frac{G^2 N \left|L\right|}{4}  \; .  \label{heavyheavy}
\eq
Thus, as expected, we have linear confinement with string tension
of $1/4$ in units of $G^2 N$.

Now, let us investigate the bound state equation for 
the general case.  A state containing two heavy sources with rest
frame separation $L$ has the form (suppressing orientation indices 
$\lambda$, $\rho$):
\begin{eqnarray}
  |\psi\rangle &=& \sum_p \frac{1}{\sqrt{N^{p+1}}}\int_0^{
 \klink< v^+ \sqrt{G^2 N} \kmax}
         dk_1 \cdots dk_p \, 
                   \psi(k_1,\ldots,k_p) \int_{-\infty}^\infty dl \,
                e^{2 i L l v^-} \nonumber \\
        & & \cdot b^\da(l-K/2)\, a^\da(k_1)\cdots a^\da(k_p)\, d^\da(-l-K/2)
\end{eqnarray}
where $\klink=\sum_i k_i$ is the total longitudinal momenta of the link
fields. The associated bound state equation has the form:
\begin{eqnarray}
P^- v^+ \psi(k_1,\ldots,k_p) &=&
      \frac{\klink}{2 v^+}\psi(k_1,\ldots) 
        \nonumber \\
    & & -\frac{G^2 N v^+}{8 \pi} \int \frac{dk^\prime}{
         \sqrt{k_1 k^\prime}} \frac{k_1+k^\prime}{(k_1-k^\prime)^2}
         \, \psi(k^\prime,k_2,\ldots) \,
           e^{i L v^- (k_1-k^\prime)}
        \nonumber \\
    & & +\frac{\tau}{8 \pi a^{D-3}} \int \frac{dk^\prime}{
         \sqrt{k_1 k^\prime}} \, \psi(k^\prime,k_2,\ldots) \,
           e^{i L v^- (k_1-k^\prime)}
        \nonumber \\
    & & +\frac{G^2 N v^+}{8 \pi} \int \frac{dk_1^\prime\, 
          dk^\prime_2}{
         \sqrt{k_1^\prime k_2^\prime}} \frac{k_2^\prime-k_1^\prime}{
         (k_1^\prime+k_2^\prime)^2}
         \, \psi(k_1^\prime,k_2^\prime,k_1,\ldots) \,
           e^{-i L v^- (k_1^\prime+k_2^\prime)}
        \nonumber \\
    & & +\frac{\tau}{8 \pi a^{D-3}} \int \frac{dk_1^\prime\, 
          dk^\prime_2}{\sqrt{k_1^\prime k_2^\prime}}
         \, \psi(k_1^\prime,k_2^\prime,k_1,\ldots) \,
           e^{-i L v^- (k_1^\prime+k_2^\prime)}
        \nonumber \\
    & & +\frac{G^2 N v^+}{8 \pi} \frac{k_2-k_1}{\sqrt{k_1 k_2}
         (k_1+k_2)^2}
         \, \psi(k_3,\ldots) \,
           e^{i L v^- (k_1+k_2)}
        \nonumber \\
    & & +\frac{\tau}{8 \pi a^{D-3}} \frac{1}{\sqrt{k_1 k_2}}
         \, \psi(k_3,\ldots) \,
           e^{i L v^- (k_1+k_2)}
        \nonumber \\
    & & + v^+ \left(\mbox{``link-link'' interactions from 
                        Eqn.~(\ref{redham})}\right)   
        \nonumber \\
    & & + \parbox[t]{3in}{``source-link'' interactions on the other end 
               of the string (complex conjugate of above).}   
              \label{longbound}
\end{eqnarray}
Note that the first term scales with momentum as $\left(\klink\right)^1$, 
the $\tau$ 
terms scale as $\left(\klink\right)^0$,
and all other terms scale as $\left(\klink\right)^{-1}$.
Thus, the longitudinal momentum scale is determined as a balance between the 
first term and the remaining terms of the Hamiltonian.
As opposed to the closed string, an instantaneous-annihilation process
via $A_+$
can act in the $p=2$ link sector and the ``pinching'' terms 
({\em exempli gratia} $\lambda_3$)
do not come into play; otherwise, there are
no changes in the ``link-link'' interactions from the closed string case.

There is a variety of possible cutoffs for $k_i$  
that one might choose.
The cutoff used here $ \klink< v^+ \sqrt{G^2 N} \kmax$ is convenient because
the ``link-link'' interactions, which act on the interior of the 
string, conserve $\klink$.
Holding the other momenta constant and ignoring particle number changing
interactions, we see that $k_1$ has the limiting behaviors\footnote{
The $\beta$ here is conventional notation for the longitudinal
endpoint behavior, not to be confused with the
plaquette coupling in (\ref{redham}).}
\begin{eqnarray}
\lim_{k_1 \to 0} \psi(k_1,\ldots) &\propto& k_{1}^{\beta} \; , 
  \hspace{.25in}0<\beta<1/2\; , \hspace{.1in} 
   \frac{\mu^2}{G^2 N} = \beta \tan(\pi \beta) \\
\lim_{k_1 \to \infty} \psi(k_1,\ldots) &\propto& 
         \left( \frac{\tau}{a^{D-3}k_1^{3/2}}-
                \frac{G^2 N v^+}{k_1^{5/2}}
   \right) e^{i L v^- k_1} \; . \label{k12}
\end{eqnarray}
The first relation is exactly the same ``endpoint behavior'' that
occurs for particles in the interior of the string. The second relation
(\ref{k12}) implies that the eigenvalues of $v^+ P^-$ converge as
$1/\left(\kmax\right)^5$ for $\tau=0$ and 
$1/\left(\kmax\right)^3$ for $\tau$ nonzero.

We solve the bound state equation (\ref{longbound}) by rescaling
and discretising the longitudinal momenta
\be
k_i \,\Rightarrow\, \frac{v^+ \sqrt{G^2 N} \kmax}{\dkmax}\, \tilde{k}_i
    \; , \;\; \tilde{k}_i\in \{1/2,3/2,\ldots\} 
                \;, \label{lnote}
\eq
where $\dkmax\ge\dklink=\sum_i \tilde{k}_i$ is the maximum total 
discretised momentum.  The ratio $\kmax/\dkmax$ parametrises
the coarseness of the discretization.
The discretised bound state equation, expressed in terms of
the dimensionless parameters $\kmax$ and $L  \sqrt{G^2 N}$,
is linear in the couplings 
\be
         \sqrt{G^2 N}\; , \hspace{0.2in}
         \frac{\mu^2}{\sqrt{G^2 N}} \; , \hspace{0.2in} 
        \frac{\lambda_i}{\sqrt{G^2 N} a^{D-2}} \; , \hspace{0.2in}
        \frac{\tau}{a^{D-3}} 
\eq
which all have units of mass.  For large $L$, the wavefunction 
becomes oscillatory.  In practice, this forces us to extract
the heavy quark potential from relatively small longitudinal separations:
of order a few transverse lattice spacings.


\section{Lorentz Invariance}
\label{lorentzinvariance}

One of the advantages of light-front quantisation is that boosts ---
both longitudinal and transverse --- become simple.  However, our
transverse lattice regulator, along with our use of effective
degrees of freedom, break transverse Lorentz covariance.  Instead,
we will demand Lorentz covariance under transverse boosts for the 
lightest glueball states; we will use this to help determine the effective 
potential.

In Section~\ref{lorentz}, we will construct states with 
nonzero transverse momentum and discuss the associated calculations
of eigenstates of $P^-$.  In Section~\ref{dispersion}, we will 
discuss  how these boosted states can be used to help determine
the effective potential. 

\subsection{Non-zero transverse momentum ${\bf P}$}
\label{lorentz}

For the sake of clarity, we will show the transverse 
coordinates $\bf x$ for the link fields $a^\da_\lambda(k, \bf x)$.
The transverse coordinates will later be dropped:  we shall
see that non-zero transverse momentum is equivalent to introducing
various phase factors in the Hamiltonian in the Eguchi-Kawai reduced theory.
In the following, we will consider vanishing winding number ${\bf n}=0$;
the extension to nonzero winding is discussed in 
Appendix~\ref{nonzerowinding}.

A generic $p$-link state of transverse momentum $P$ 
has the form:
\be
   \left | \Psi(P^+,{\bf P})\right\rangle =
          \frac{1}{\sqrt{N^p}} \sum_{\bf y}
        {\rm e}^{i \bf{P}\cdot \left({\bf y}+{\bf \bar x}\right)} 
      \Tr\left\{ a_{\lambda_1}^\da(k_1,{\bf x}_1+{\bf y}) \, 
        a_{\lambda_2}^\da(k_2,{\bf x}_2+{\bf y})
        \cdots  a_{\lambda_p}^\da(k_p,{\bf x}_p+{\bf y})\right\} 
        \left|0\right\rangle
        \; .  \label{dfunct}
\eq
where the winding number vanishes $\sum_{i=1}^p \widehat{\lambda}_i = 0$
and ${\bf y}$ is summed over links of the transverse lattice.
The transverse coordinates must obey the formula 
\be
{\bf x}_i = {\bf x}_{i-1} +a \,\widehat{\lambda}_{i-1} \;, 
        \;\;\;\; i<1\le p\; .
\eq
Next, we must choose a phase convention for the overall
phase ${\bf \bar x}$ as a function of ${\bf x}_1,{\bf x}_2,\ldots$.
We have found a longitudinal momentum weighted center of mass
\be
        {\bf \bar x} = \frac{1}{P^+}\sum_{i=1}^p k_i \left(x_i+
                   \frac{a \, \widehat{\lambda}_i}{2}\right)
               \label{com}
\eq
to be especially convenient.  For this choice of ${\bf \bar x}$,
the component $M_{-r}$ of the angular momentum--boost tensor $M_{\mu \nu}$,
which generates  Lorentz boosts in direction $r$ and obeys
\be
        [P^-, M_{-r}]= {\rm i} P_{r} \; , \;\; [P^+, M_{-r}] = 0 \; , \;\;
        [P^{r},M_{-s}] = -{\rm i} g^{r}_{s} P^+ \ ,
\eq
takes on a particularly simple form:
\begin{eqnarray}
{\rm e}^{-{\rm i}b^r M_{-r}} \, a_\lambda^\da(k, {\bf x}) \, 
{\rm e}^{{\rm i}b^r M_{-r}} & = & a_\lambda^\da(k, {\bf x})\,
{\rm e}^{-{\rm i}k {\bf b}\cdot {\bf x}}  \\
{\rm e}^{-{\rm i}b^r M_{-r}}\, |\Psi(P^+, {\bf P})\rangle & = & 
|\Psi(P^+, {\bf P}- {\bf b} P^+)\rangle \; .
\end{eqnarray}
In the computation of large-$N$ matrix elements 
of an operator ${\cal O}$
\be
        \langle \Psi^\prime(P^+,{\bf P})|\,{\cal O}\,|\Psi(P^+,{\bf P})\rangle
        =\langle \Psi^\prime(P^+,0)|\,{\rm e}^{{\rm i}P^r M_{-r}/P^+}\,{\cal O}
        \,{\rm e}^{-{\rm i}P^r M_{-r}/P^+}\,|\Psi(P^+,0)\rangle \; ,
\eq
one obtains phase factors in the Eguchi-Kawai reduced theory.  
These phase factors
correspond to any shift in the weighted center of mass (\ref{com}).
For example, the interaction
\be
 {\cal O} = \sum_{\bf x} \Tr\left\{
        a^\da_{-\lambda}(k_1,{\bf x})\,
        a^\da_{\lambda}(k_2,{\bf x}-a \widehat{\lambda})\,
        a^\da_{\rho}(k_3,{\bf x}-a \widehat{\rho})\,
        a^\da_{-\rho}(k_4,{\bf x})\right\} \delta(k_1+k_2-k_3-k_4)
\eq
picks up the phase factor
\be
        {\rm e}^{{\rm i}P^r M_{-r}/P^+}\,{\cal O}\,
        {\rm e}^{-{\rm i}P^r M_{-r}/P^+}
        = {\cal O} \, \exp\!\left(-{\rm i}\,\frac{a (k_1+k_2)}{2 P^+}\,
        {\bf P}\cdot\left(\widehat{\lambda}-\widehat{\rho}\right)\right) 
        \; . \label{pexam}
\eq
We can see from this example that the phase factors
are functions only of longitudinal momenta and orientation
indices $\lambda$ and $\rho$.  Thus, Eguchi-Kawai reduction of
the theory remains.

Although we have constructed states of definite transverse momentum,
the construction of the operator ${\bf P}$ itself is somewhat 
problematic.  Consider the most natural definition,
\be
        {\bf P}=\frac{i}{2 a} \sum_{{\bf x},\lambda} \int dk\, 
        \left[a_\lambda^\da(k,{\bf x}+a \hat{r})-a_\lambda^\da(k,{\bf x}-a 
        \hat{r}) \right]  a_\lambda(k,{\bf x}) \; ,
\eq
which has the correct classical continuum limit.  When we apply this operator
to a closed colour loop, for example (\ref{typ}), we obtain a state that is
no longer a closed colour loop and whose energy therefore diverges.
Evidently, the correct form of ${\bf P}$ is quite complicated.

In numerical work, it is advantageous to work with a real-valued
representation of the Hamiltonian.  We note that, in the 
$x^r$ direction, the boost operator
$M_{-r}$ breaks the transverse reflection symmetry ${\cal P}^r:x^r \to -x^r$ 
of the theory.  The boost changes sign under transverse reflections,
\be
        {\cal P}^r \, {\rm e}^{{\rm i}b M_{-r}} \,{\cal P}^r = 
        {\rm e}^{-{\rm i}b M_{-r}}\;\;\;\;\mbox{(no sum over $r$)} \; .
\eq
For a boost in the $x^r$ direction, 
we introduce a basis with states that are even and 
odd under ${\cal P}^r$.
We use states of the form:
\begin{eqnarray}
        |\psi A\rangle &=& \cos\left(\frac{P^r M_{-r}}{P^+}\right)
        \left(1+{\cal P}^r\right)
        |\Psi(P^+,0)\rangle \;\;\;\;\mbox{(no sum over $r$)} \\
        |\psi B\rangle &=& \sin\left(\frac{P^r M_{-r}}{P^+}\right)
        \left(1-{\cal P}^r\right)
              |\Psi(P^+,0)\rangle \;\;\;\;\mbox{(no sum over $r$)}
\end{eqnarray}
and the associated phase factors (\ref{pexam}) 
become sines and cosines.  The states $|\Psi(P^+,0)\rangle$
are constructed to have definite charge conjugation 
${\cal C}\,|\Psi(P^+,0)\rangle=\pm\,|\Psi(P^+,0)\rangle$.

\subsection{Dispersion Relations.}
\label{dispersion}

We now indicate how a  scaling analysis can be carried out,
by enforcing $SO(3,1)$ 
Lorentz invariance of physical observables. There are a number
of relativistic criterion which one might apply. 
With only the spectrum of
glueball mass ratios, as a function of the couplings in $V[M]$, one might
try to obtain  Lorentz multiplets. This is a rather narrow test
however, since it pays no attention to scalar boundstates. A better measure
of relativistic covariance is the glueball dispersion formula. 
Perfectly relativistic glueballs should satisfy
\be
2P^+ P^-  = {\cal M} ^2 + |{\bf P}|^2 \label{shell} \;.
\eq
Transverse lattice glueballs will in general satisfy
\be
      2P^+ P^- =  G^2 N \left( \dm^{2}_{0} + \dm_{1}^{2}\, a^2
      |{\bf P}|^2 + \dm_{2}^{2}\, a^4 |{\bf P}|^4 + 
             \cdots \right) \label{latshell}\; .
\eq
We have used $G^2 N$ as 
the overall scale, so that the measurable coefficients 
${\cal M}_{i}^{2}$ above depend only upon the dimensionless 
ratios 
\be
        m^2 = {\mu^2 \over G^2 N}  \; , \;\;\;\; 
        \newl_i = {\lambda_i \over a^{D-2} G^2 N}  \; , \;\;\;\; 
        \newtau = \frac{\tau}{\sqrt{G^2 N} a^{D-3}} \; , \;\;\;\;
         b = {\beta \over a^{D-2} G^2 N} \; .
\eq
In order that (\ref{latshell}) agree with (\ref{shell})
one can tune the couplings so that 
$\dm_{2}^2= 0$ and $a^2 G^2 N \dm_{1}^{2} = 1$
for each glueball in the low-lying spectrum. 
Due to the fact that
there are transverse lattice hopping terms in $P^-$, even when $V=0$, the
dispersion relation automatically becomes asymptotically quadratic
for small $a {\bf P}$. For this reason, the condition $\dm_{2}^2=0$ 
is more difficult to use to fix $V$. 
The condition $a^2 G^2 N \dm_{1}^{2} = 1$, 
which is more useful,
ensures that the speed of light is 
isotropic. The rotational invariance of the speed of light
in the transverse directions was discussed in Ref.~\cite{bard2}.
In order to ensure
that the  speed of light in transverse
directions $c_T$ equals that in the $x^{D-1}$ direction $c_L$ 
(set to one by convention),
one must determine the dimensionless combination $a^2 G^2 N$.

One way is to measure both the longitudinal and transverse string tensions. 
This also allows one to express the glueball
spectrum in terms of the string tension which is essential for making an
absolute prediction of the glueball masses.
Suppose we make the following three measurements of dimensionful quantities
\begin{eqnarray}
{v_1 G^2 N} & = & c \sigma \label{mass} \\
{v_2 G^2 N} & = & a^2 \sigma^2 \label{tten} \\
{v_3 G^2 N} & = & \sigma \label{longten}
\end{eqnarray}
where $v_1$, $v_2$, $v_3$ are measured dimensionless numbers, $c$ is an
unknown constant, and  $\sigma$ is the continuum string tension.
Eqn.~(\ref{mass}) comes from a glueball mass measurement. 
Eqn.~(\ref{tten}) comes from  the transverse winding
spectrum (Section~\ref{trans}). Eqn.~(\ref{longten}) comes from  
the asymptotic 
heavy-source potential in the $x^{D-1}$ direction (Section~
\ref{long}).
We have also
demanded that the longitudinal and transverse tensions are equal
$\sit = \sil = \sigma$. 
Scaling means that the right hand sides of the above equations
are not multiplied by dimensionless functions of $a^2 G^2 N$. 
We now show that this is enough to solve the theory completely. 

We use the slope of the dispersion relations at ${\bf P} = 0$ for a few
low-lying glueballs
to estimate the  scaling trajectory of $V$, by demanding isotropy of the 
speed of light. 
{}From (\ref{shell}) and (\ref{latshell}) one needs to determine 
$G^2 N a^2$ in order to compare transverse and longitudinal scales when
comparing the speeds of light.  
{}From (\ref{tten}) and (\ref{longten}) we can deduce 
\be
G^2 N a^2 = v_2 / v_{3}^{2}
\eq
There are a number of self-consistency
checks one can perform, such as examining deviations from quadratic
dispersion (non-zero $\dm_{2}^{2}$), the Lorentz multiplet structure in the
glueball spectrum, and rotational invariance of the heavy source potential.
Similarly, we can now measure the lattice spacing $a$ and glueball masses $c$
in units of $\sigma$.


\section{Calculations}
\label{calculation}

We now implement the procedures described in the first half of the paper
for the case of $2+1$-dimensional pure large-$N$ gauge theory. The main
interest firstly is to compute glueball masses in physical units. These
can be compared with results from established techniques to ascertain whether
the transverse lattice theory produces correct results, at least to the
level of approximation we use. 
However, it is worth bearing in mind that, apart from
confirming known data, the light-front
Hamiltonian approach is uniquely suited to the calculation of hadronic
wavefunctions in a general Lorentz frame, and it is the latter which 
chiefly motivates us. We will exhibit the structure of these wavefunctions
also. We have also begun analogous calculations in $3+1$ dimensions, which
are briefly discussed in the conclusions.

\subsection{2+1 Dimensions}
\label{twoplusone}

Yang-Mills theory in 2+1 dimensions is superrenormalisable. Rather than
a logarithmic Coulomb potential, a linearly confining interaction 
appears to be dynamically generated \cite{teper}. Therefore  it is rather
similar to $3+1$-dimensional QCD. It provides
a useful test-bed for non-perturbative calculations, given that two
space dimensions can be more efficiently handled numerically than three.
The discussion of the first half of the paper can be trivially modified
for $2+1$ dimensions 
(the reader is referred to \firstpaper{} for further detail). 
Our light-front Hamiltonian (\ref{redham}) reduces in $2+1$ dimensions
to
\begin{eqnarray}
 P^-  & = & \int dx^- 
         - {G^2 \over 4} \Tr\left\{  J^{+} 
        \frac{1}{\partial_{-}^{2}} J^{+} \right\} 
        + \mu^2  \Tr\left\{MM^{\da}\right\} \nonumber \\ 
        && + {\lambda_1 \over a N}
        \Tr\left\{ M M^{\da}MM^{\da} \right\} + {\lambda_2 \over a N}  
        \Tr\left\{ MMM^{\da} M^{\da} \right\} +  {\lambda_3 \over a N^2} 
        \left( \Tr\left\{ MM^{\da} \right\} \right)^2 
                        \\
        J^{+} &=&  i M \stackrel{\leftrightarrow}{\partial}_{-} 
                M^{\da}  + M^{\da} \stackrel{\leftrightarrow}{\partial}_{-} M
\end{eqnarray}
where now $a G^2 \to g^{2}_{3}$ as $a \to 0$, with $g^{2}_{3}$ the standard
2+1 dimensional Yang-Mills coupling in the continuum 
limit.\footnote{In \firstpaper{}, we did not
clearly distinguish between the $a$-dependent transverse
lattice coupling $G$ and the continuum coupling. 
Moreover, the normalisation of the continuum gauge
coupling in that work was nonstandard. Replace $G^2$ by $2 g^2$ 
and $m^2$ by $m^2/2$ to convert expressions in this paper 
to those appearing in \firstpaper{}.} 
Otherwise the analysis is
unmodified (in particular, the scaling analysis of Sec.~\ref{dispersion}).
The link field $M(x^-)$, of course, now points in one transverse direction
only. Under large-$N$ reduction it depends only upon $x^-$, and is equivalent
to two real adjoint matrix fields. We are now ready to diagonalise $P^-$
in the Fock space of $M({\bf x})$, as a function of 
$m, l_1, l_2, l_3$. 
$G^2 N$ will be used to set the overall scale for dimensionful
physical quantities. It will be determined in terms of a physical 
scale by measurements of the string tension. The glueball spectrum
is classified by $|{\cal J}|^{{\cal P}_{1} {\cal C}}$, for
angular momentum ${\cal J}$, $x^1$ reflection symmetry ${\cal P}_{1}$,
and charge conjugation ${\cal C}$. The latter two are exact while first
can only be estimated approximately on the transverse lattice.
We follow the computational methods established in Refs.~\cite{model}
for adjoint scalar matter in two-dimensions, together with various 
improvements described in the Appendices of \firstpaper{} and
in the next section.

\subsection{Methods}
\label{methods}

Our basic numerical method is DLCQ \cite{dlcq}, 
where we discretise the longitudinal 
momentum $P^+ = 2 \pi K / {\cal L}$ into odd half-integers
in the bound state equation.  
$K$ measures the coarseness
of the discretisation and $K\to\infty$ is the longitudinal continuum.
This approach has several advantages for numerical work:
the resulting Hamiltonian matrix is sparse and matrix elements
of the interaction are easy to compute.  The main disadvantage of DLCQ is
that numerical convergence tends to be slow.  In Appendix~C of 
\firstpaper{}, we introduced an approach which eliminated the most
severe convergence problems, the $1/K^{2 \beta}$ and $1/\sqrt{K}$ errors,
while minimising the remaining $1/K$ error associated with the 
instantaneous interactions.  We have made some subsequent 
refinements to the method which are described in Appendix~\ref{appendixc}.
%
%
%
\begin{figure}
\centering
\begin{tabular}{c@{}c}
$\displaystyle\frac{{\cal M}^2}{G^2 N}$ &
\BoxedEPSF{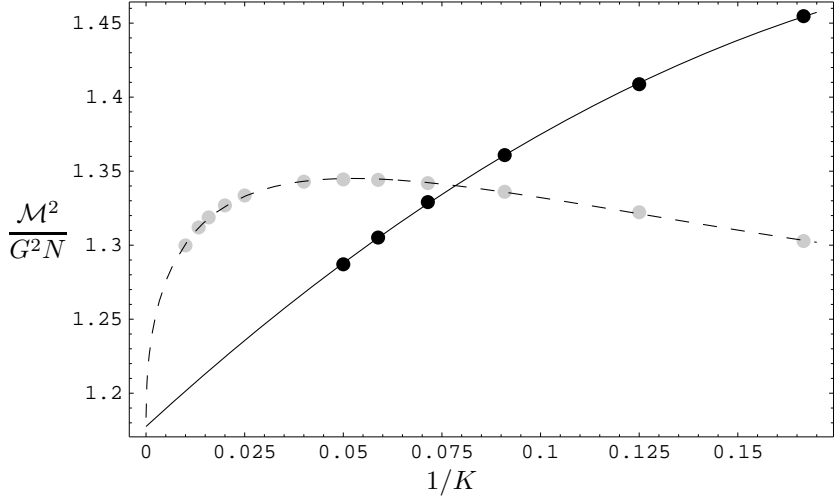 scaled 800}\\
&$1/K$
\end{tabular}
\caption{
Convergence of the lowest glueball mass as a
function of $1/K$ for a 4-link truncation
and $\mu^2=\lambda_i=0$.
The gray points are from conventional DLCQ and
the dashed line is a fit to $
1.1834 + 1.6395/\sqrt{K} - 5.0731/K + 4.3513/K^{3/2}$.
The black points are from a calculation using improved matrix
elements and are fit to
$1.17752+2.44169/K- 4.68183/K^2$.
Thus, given data points in the range $10<K<20$,
the improved matrix elements are necessary for a
reliable extrapolation.
\label{kconverge}}
\end{figure}
Because of these improvements, we are able to extrapolate
to infinite $K$ with some confidence.  For instance, in 
Figure~\ref{kconverge} we plot 
the eigenvalue of the lowest glueball, calculated using 
improved matrix elements, as a function of $1/K$.  
It is clear that a fit to $c_0+c_1/K$ should work quite well for $K$
between 10 and 20.  

In additional to the harmonic resolution $K$, we truncate
the Fock space in the number of links $p$.  We calculate the convergence
in $p$-truncation numerically by looking at the
convergence of the lowest eigenvalue for winding number $|{\bf n}|=1$
and $
(K,p\mbox{-truncation})=(19/2,3)$,  $(25/2,3)$, 
  $(39/2,3)$,  $(19/2,5)$,  $(25/2,5)$,  $(23/2,7)$.
Then we fit the eigenvalue to the function
\be
       c_0 + \left(c_1+\frac{c_2}{K}\right) {\rm e}^{c_p p}+
        \frac{c_3}{K} \; .
\eq
It is satisfying that the results for $c_p$, typically of
order -1, agree with the analytic estimate
from Appendix~B of \firstpaper{}.  We assume that, for sufficiently
large $p$-truncation, $c_p$ is a universal constant; it can
be equally well applied to the spectra, nonzero winding $|{\bf n}|$,
and the heavy source potential.

Using this result for $c_p$, we calculate spectra for various
$(K,p\mbox{-truncation})=(18/2,6)$,  $(18/2,8)$, 
  $(20/2,6)$,  $(20/2,8)$,  $(24/2,6)$,  $(32/2,6)$ and
extrapolate to the continuum using the function
\be
       c_0 + c_1 {\rm e}^{c_p p}+\frac{c_2}{K} \; .\label{fit11}
\eq
We use the same extrapolation procedure for nonzero winding 
$|{\bf n}| \neq 0$.  In this case we use
$(|{\bf  n}|,K,p\mbox{-truncation})=( 2,20/2,4)$  $( 2,20/2,6)$, 
  $( 2,24/2,4)$,  $( 2,28/2,4)$,  $( 3,21/2,5)$,  $( 3,21/2,7)$,  
  $( 3,23/2,5)$,  $( 3,27/2,5)$,  $( 4,20/2,6)$,  $( 4,20/2,8)$,  
  $( 4,22/2,6)$,  $( 4,26/2,6)$.
After extrapolating in $K$ and $p$-truncation using Eqn.~(\ref{fit11}),
we fit the lowest eigenvalue to the form 
$(\sit a |{\bf n}|)^2/(G^2 N)+c_1+c_2/|{\bf n}|^2$ as 
discussed in \firstpaper{}.

\begin{figure}
\centering
\begin{tabular}{@{}c@{}c@{}}
$\displaystyle\frac{v^+ P^-}{\sqrt{\sigma}}$ &
\BoxedEPSF{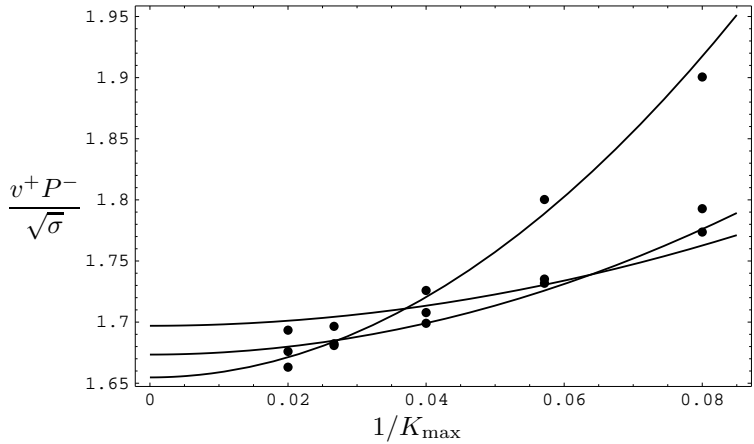 scaled 700}\\
&$\displaystyle 1/\dkmax$
\end{tabular}
\caption{
Convergence of the $ |{\bf n}|=0$ heavy source potential with $1/\dkmax$
for $\kmax=4$, 5, 8 and a 2-link truncation.
The couplings are from the $m=0.180255$ datum of Table~\ref{tabtraj1} 
and the longitudinal separation is 
$L =4/\sqrt{G^2 N}=1.750/\sqrt{\sigma}$.
The fit function is from Eqn.~(\ref{lfit4}).
\label{lconverge1}}
\end{figure}
\begin{figure}
\centering
\begin{tabular}{@{}c@{}c@{}}
$\displaystyle\frac{v^+ P^-}{\sqrt{\sigma}}$ &
\BoxedEPSF{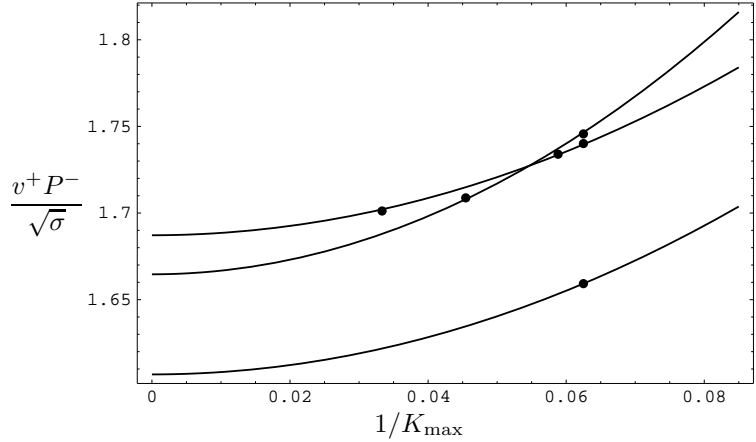 scaled 700}\\
&$\displaystyle 1/\dkmax$
\end{tabular}
\caption{
Convergence of the  $ |{\bf n}|=0$ 
heavy source potential with $1/\dkmax$ for various
$\kmax$ and $p$-truncations as given by (\ref{lfit3}).
The couplings are from the $m=0.180255$ datum of Table~\ref{tabtraj1}
and $L =4/\sqrt{G^2 N}=1.750/\sqrt{\sigma}$.
The fit function is from Eqn.~(\ref{lfit2}).
\label{lconverge2}}
\end{figure}

The most tricky calculation is the heavy source
potential.  Here we have to extrapolate in $\dkmax$,
$\kmax$, and $p$-truncation.  In addition, we must
choose reasonable values for the longitudinal separation
$L$.  For example, in Fig.~\ref{lconverge1}, we see that the eigenvalues
are well fit to the function
\be
 \frac{v^+ P^-}{\sqrt{\sigma}} = 1.64873 + \frac{3.08839}{\left(\kmax\right)^3}
            + 0.64118 \left(\frac{\kmax}{\dkmax}\right)^2 \label{lfit4}
\eq
with an RMS deviation of 0.0088.  Although we would expect a the 
leading finite $\dkmax$ error to be of order $\left(\kmax/\dkmax\right)^1$,
we find the quadratic term to be much more important.  
Thus, we shall use the fitting function
\be
 \frac{v^+ P^-}{\sqrt{G^2 N}} = c_0 + \frac{c_1}{\left(\kmax\right)^3}
            + c_2 \left(\frac{\kmax}{\dkmax}\right)^2+c_3\, {\rm e}^{c_p p}
        \label{lfit1} \; .
\eq
In subsequent calculations, we fit to (\ref{lfit1}) using the values
\begin{eqnarray}
  \lefteqn{
  \left( |{\bf n}|,\dkmax,\,
p\mbox{-truncation},\kmax\right)=
 (0,32/2,2,4),\; (0,32/2,4,4),}
       \nonumber\\
  &&  (0,32/2,2,5),\;(0,34/2,2,4),\;(0,44/2,2,5),\;(0,60/2,2,4),\;(1,19/2,3,4),
        \nonumber\\ 
  &&  (1,19/2,5,4),\;(1,19/2,3,5),\;(1,33/2,3,4),\;(1,33/2,3,5),\;(1,49/2,3,4)
          \; .\label{lfit3}
\end{eqnarray}
Note that for the 
heavy source calculation, the conserved quantity ${\bf n}$, analogous to
winding number, is the number of links minus number of
anti-links, {\em id est} 
the transverse separation of the sources in lattice units.
Example data points are shown in Fig.~\ref{lconverge2} along with the
associated fit function
%
%
%
\be
 \frac{v^+ P^-}{\sqrt{\sigma}} = 1.54997 + \frac{2.95576}{\left(\kmax\right)^3}
            +0.838171 
\left(\frac{\kmax}{\dkmax}\right)^2+0.773708\, {\rm e}^{c_p p} \; .
        \label{lfit2}
\eq
In order to measure the heavy source potential and test its rotational 
invariance, we calculate the potential for  $|{\bf n}|=0$
and $L\sqrt{G^2 N}=3$, 4, and 6. Then we fit to the functional
form~\cite{lsw}
\be
   \frac{v^+ P^-}{\sqrt{G^2 N}} = 
	c_0\, L\sqrt{G^2 N}+c_1+\frac{c_2}{L \sqrt{G^2 N}} 
		\; . \label{longpot}
\eq
Next we calculate the potential for $|{\bf  n}|=1$ and $L\sqrt{G^2 N}=3$; 
we then demand rotational invariance for this datum.


\section{Results}
\label{results}

\subsection{First Principles}

To search for a scaling trajectory that restores Lorentz covariance,
we applied a $\chi^2$ test for certain observables. First, we 
investigated which observables would be useful in this respect. The 
natural candidates are the $c_{T}/c_{L}$ ratios deduced from glueball
dispersion relations, the Lorentz multiplet degeneracy in the spectrum,
and rotational invariance of the heavy source potential.

In order to determine which glueballs it makes sense to include in the 
$c_{T}/c_{L}$ part of the  $\chi^2$ test, 
for each low-lying glueball we determined the $\newl_i$ trajectory such that
$c_{T}/c_{L} = 1$ for that glueball. In the low-lying spectrum, only
the result for the first excited state $0^{++}_{*}$ was markedly incompatible
with the rest. In fact, it is consistently imaginary. We show below
why this artifact is to be 
expected of our truncation of the effective potential
to fourth order. 
Therefore, excluding this state,
we included the $c_{T}/c_{L}$ ratio of the
lowest seven glueballs: $0^{++}, 2^{++}, 2^{-+}, 0^{--}, 0^{--}_{*}, 2^{+-},
2^{--}$. These are the lowest seven glueballs for a broad range of $\newl_i$
at fixed $m$, and also the seven lowest in the ELMC simulations \cite{teper}.

%
%
%
\begin{table}
\centering\[
\renewcommand{\arraystretch}{1.25}
\begin{array}{c|ccccc|ccc}
 m & \newl_1 & \newl_2 & \newl_3 & \newtau & 
 \displaystyle\frac{G^2 N}{\sigma} &\chi^2 & c_p & 
 \displaystyle\frac{a^2 \sit^2}{G^2 N} 
\\[7.5pt]\hline\hline
0.134186 & -0.1263 & -0.1178 & 3.5004 & -0.3289 & 5.3375 & 17.5267 & -1.1680 
& 0.2834\\
0.180255 & -0.0772 & -0.2508 & 221.03 & -0.9606 & 5.2220 & 12.1472 & -1.0791 
& 0.2699\\
0.227546 & -0.1606 & -0.1754 & 7.3792 & -1.1712 & 5.4358 & 13.6865 & -1.1836 
& 0.3473\\
0.276458 & -0.1993 & -0.2049 & 6.4575 & -0.9429 & 5.0986 & 10.8355 & -1.1455 
& 0.3860\\
0.327475 & -0.1783 & -0.2133 & 9.4007 & -0.8262 & 4.6138 & 12.6389 & -1.4465 
& 0.4592\\
0.381194 & -0.2032 & -0.1397 & 3.4963 & -1.0544 & 4.5280 & 20.9469 & -1.7720 
& 0.5671\\
0.438370 & -0.1642 & -0.1254 & 2.0687 & -1.2205 & 4.3206 & 21.8354 & -2.2138 
& 0.6670\\
0.500000 & -0.1731 & -0.1284 & 2.3097 & -1.4091 & 4.1691 & 22.6331 & -2.2572 
& 0.7603
\end{array}\]
\caption{The scaling trajectory which minimises the $\chi^2$ test of
Lorentz covariance. The couplings $\newl_i$, $\newtau$, and the
overall scale $G^2 N/\sigma$ are all obtained from the test. $m$ is 
equivalent to the lattice spacing degree of freedom.
\label{tabtraj1}}
\end{table}

\begin{figure}
\centering
\BoxedEPSF{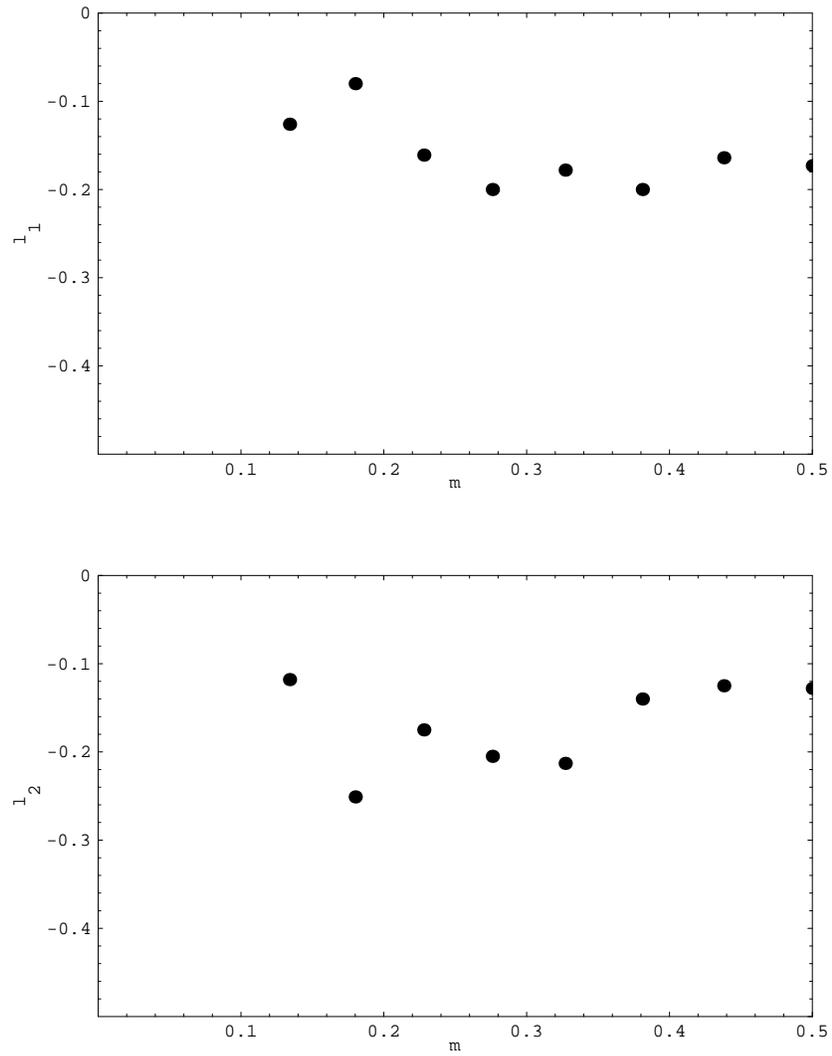 scaled 700}
\caption{The trajectory which minimises the $\chi^2$ test of Lorentz
covariance; see Table~\ref{tabtraj1}. 
\label{traj1}}
\end{figure}

Another possible test of Lorentz covariance is the degeneracy
of the parity doublets for the $2^{\pm+}$ and the $2^{\pm-}$. 
However we find that there is no point in our parameter
space where the $2^{++}$ and the $2^{-+}$ are degenerate and we
do not include this pair in our $\chi^2$ test.
For the $2^{+-}$ and $2^{--}$, we have the opposite problem:
the two levels are almost degenerate for a vary wide range of
couplings.  As a consequence, it does not provide a very useful
criterion for finding the correct scaling trajectory (although
it was included).

In addition, we demand rotational invariance of the heavy source potential
based on measurements at $|{\bf n}|=0$ and 1.
As described in subsection~\ref{dispersion}, the longitudinal string tension 
measurement determines the overall
scale $G^2 N/\sigma$ and, as with our other measurements, we assume that
there is some error associated with this. In principle, 
one might infer
the overall scale from demanding consistent dispersion relations of the 
glueballs,
but we found this method to give unreliable results.

The coupling trajectory we find from the resulting $\chi^2$ test is
shown in Table~\ref{tabtraj1} and Figure~\ref{traj1}. 
As previously found 
in \firstpaper{}, 
the magnitude of $\newl_3$ always comes out much larger than that of 
$\newl_1$ and $\newl_2$,
essentially infinite. Here we are able to confirm that this is required
by Lorentz covariance. We also elaborate on this below.
Although there are some fluctuations in the coupling trajectory as $m$ is
varied, it is nevertheless remarkably close to the trajectory found
in \firstpaper{} on the basis of a {\em fit} to the ELMC spectrum.

\begin{figure}
\centering
\begin{tabular}{c@{}c}
$\displaystyle\frac{v^+ P^-}{\sqrt{\sigma}}$ &
\BoxedEPSF{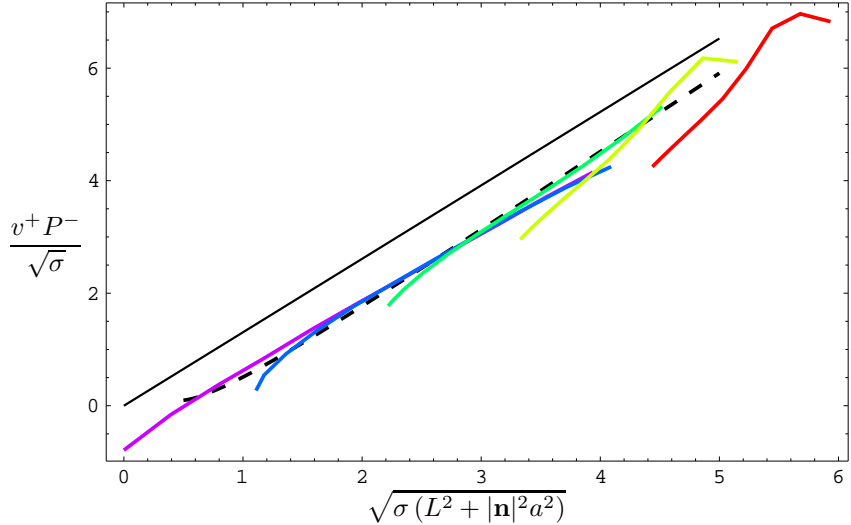 scaled 800}\\
&$\displaystyle\sqrt{\sigma\left(L^2+|{\bf n}|^2 a^2\right)}$
\end{tabular}
\caption{
The shaded lines represent the  heavy source 
potential as a function of 
separation $\sqrt{\sigma\left(L^2+ |{\bf n}|^2 a^2\right)}$ 
for $ |{\bf n}|=0,$ 1, 2, 3, 4.
The dashed line is from Eqn.~(\ref{longpot}) 
and the solid line is the heavy-heavy Coulomb potential
from Eqn.~(\ref{heavyheavy}).
The eigenvalues were obtained by extrapolating in 
$\dkmax$, $\kmax$, and $p$-truncation; the 
couplings are from the $m=0.180255$ datum of Table~\ref{tabtraj1}.
\label{upotential}}
\end{figure}

In determining the trajectory, we inferred roundness of the heavy source
potential from measurements at $|{\bf n}|=0$ and 1.  If we plot
the heavy source potential for more values of $|{\bf  n}|$, 
we observe that rotational
invariance of the potential is maintained; see Fig~\ref{upotential}.  
A potential that is ``oval''--- that is, rotationally
invariant to within an overall $\sit/\sil$ --- seems
to be generic for the transverse lattice~\cite{burkardt}.  
Consequently, the crucial test is whether one obtains
consistent longitudinal versus transverse scales $\sit=\sil=\sigma$.

\begin{figure}
\centering
\BoxedEPSF{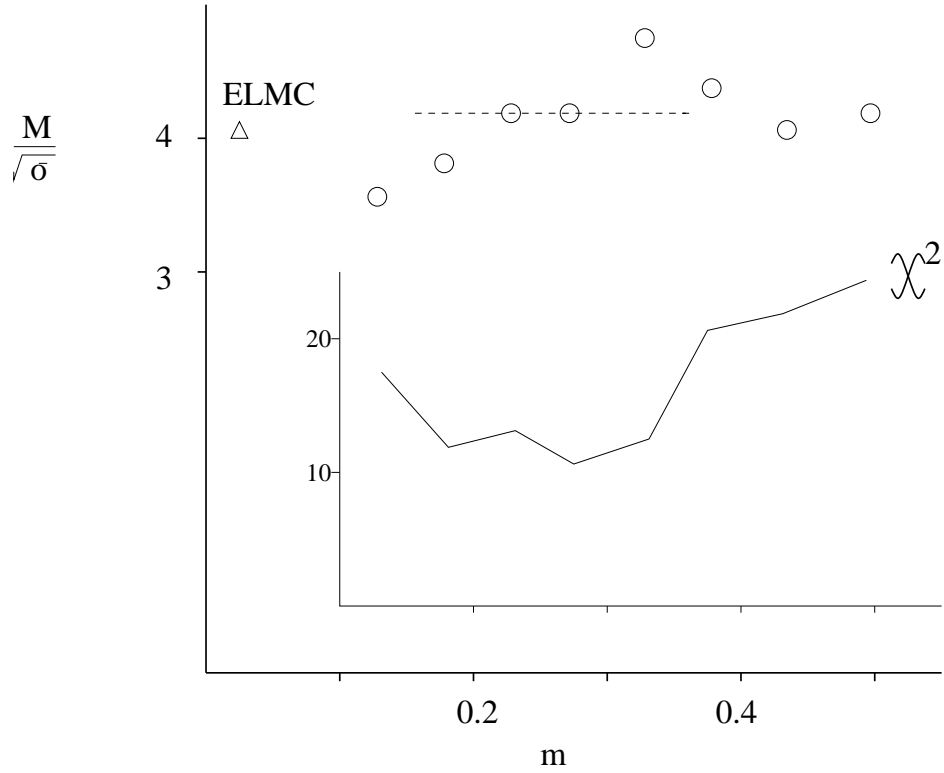 scaled 700}
\caption{The $0^{++}$ glueball mass and $\chi^2$ values
along the coupling trajectory
of Table~\ref{tabtraj1}. The size of the data points gives an indication of the
error from extrapolation to the longitudinal continuum limit, {\em id est} 
the dominant error is from scaling violation on the transverse lattice.
\label{grou}}
\end{figure}

Examining the glueball masses on this trajectory more closely, we find
results for the $0^{++}$ groundstate mass shown on Figure~\ref{grou}. 
It is to be
anticipated that the $\chi^2$ becomes poor at large and small $m$. 
At large $m$, corresponding
to large $a$ (see Fig.~\ref{spacefig}), 
the truncated effective potential cannot cope
with the transverse discretisation errors. At small $m$, the wavefunctions
are most sensitive to the the small $k^+$ region, and hence the 
longitudinal discretisation errors. It is also the edge of the 
colour-dielectric regime. The $0^{++}$ $c_{T}/c_{L}$ 
ratio is $1 \pm 0.15$ all along the coupling trajectory, and the mass scales
to some degree. Based on this, with a generous error for scaling violation,
we estimate $M_{0^{++}}/\sqrt{\sigma} = 4.05 \pm 0.3$,
in good agreement with the ELMC value\footnote{The only other accurate
value we know of comes from an SU(3) equal-time Hamiltonian lattice calculation
\cite{luo}. However, apart from $1/N^2$ corrections,
it cannot be compared directly with our result
since the authors do not measure the string tension, but give results
in terms of a  gauge coupling.}
 $4.08(7)(stat)$.

\begin{table}
\begin{center}
\renewcommand{\arraystretch}{1.25}
\begin{tabular}{|c|c|c|c|} \hline
$|{\cal J}|^{{\cal P}_{1} C}$ & $M/\sqrt{\sigma}$ &  $c_{T}/c_{L}$  \\ \hline
$0^{++}$ & $4.16 $ & $1.10$ \\ \hline
$0^{--}$ & $4.86$ & $0.71$ \\ \hline
$2^{++}$ & $5.27$ & $0.59$   \\ \hline
$2^{-+}$ & $6.76$ & $1.00$  \\ \hline
$0^{++}_{*}$ & $5.39$ & $0.96 {\rm i}$  \\ \hline
$0^{--}_{*}$ & $6.86$ & $0.87$  \\ \hline
$2^{+-}$ & $7.32$ & $0.39$  \\ \hline
$2^{--}$ & $7.64$ & $0.38$ \\ \hline
\end{tabular}
\end{center}
\caption{The glueball masses at $m=0.276$ on the scaling trajectory of
Table~\ref{tabtraj1} at the minimum  $\chi^2 = 10.84$. 
\label{table1}
}
\end{table}

At the best
$\chi^2$, we list the masses and $c_{T}/c_{L}$ ratios of the other
low-lying glueballs in Table~\ref{table1}. The level ordering of the seven
glueballs is in agreement
with the ELMC result, and the numerical values are generally not far off. 
We note however that, with one
exception, we cannot get the higher glueballs, 
along with the string tension and $0^{++}$,
to be especially 
rotationally invariant for our approximation to the 
effective potential. It is perhaps not coincidental that both the masses
and $c_{T}/c_{L}$ ratios of the higher glueballs 
come out consistently too low. The one exception,
$2^{-+}$, is in good agreement with the ELMC mass $6.89(21)(stat)$. 

\begin{figure}
\centering
\begin{tabular}{c@{}c}
$\displaystyle\sqrt{\frac{2 P^+ P^- -{\cal M}^2}{\sigma}}$ &
\BoxedEPSF{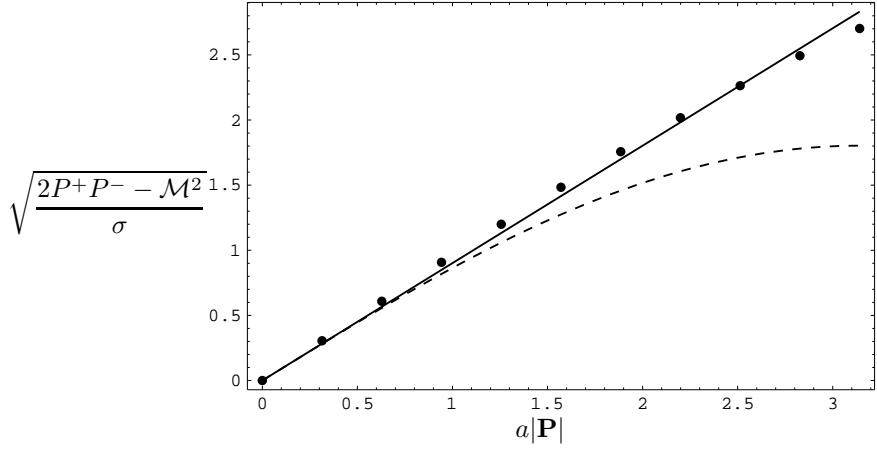 scaled 700}\\
&$\displaystyle a |{\bf P}|$
\end{tabular}
\caption{
Dispersion relation for the $0^{++}$ up to the edge of the Brillouin zone.
The solid line is the relativistically correct dispersion relation
and the dotted line is a lattice dispersion relation.
The couplings are from the $m=0.180255$ datum of Table~\ref{tabtraj1}.
\label{disp1}}
\end{figure}

\begin{figure}
\centering
\begin{tabular}{c@{}c}
$\displaystyle\sqrt{\frac{2 P^+ P^- -{\cal M}^2}{\sigma}}$ &
\BoxedEPSF{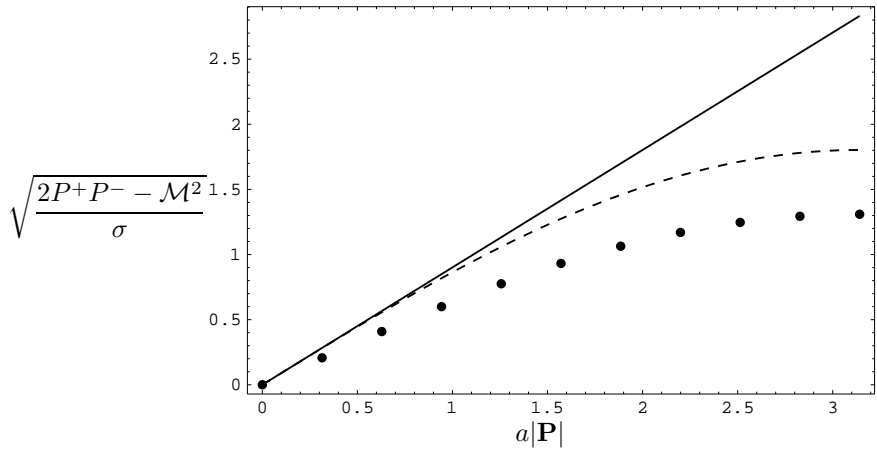 scaled 700}\\
&$\displaystyle a |{\bf P}|$
\end{tabular}
\caption{
Dispersion relation for the $0^{--}$ up to the edge of the Brillouin zone.
The solid line is the relativistically correct dispersion relation
and the dotted line is a lattice dispersion relation (normalised to the
same slope at $|{\bf P}| = 0$).  The $0^{--}$
typically has a poor slope at $|{\bf P}| = 0$ and
follows a lattice dispersion more closely.
The couplings are from the $m=0.180255$ datum of Table~\ref{tabtraj1}.
\label{disp2}}
\end{figure}

\begin{figure}
\centering
\begin{tabular}{c@{}c}
$a \sqrt{\sigma}$&\BoxedEPSF{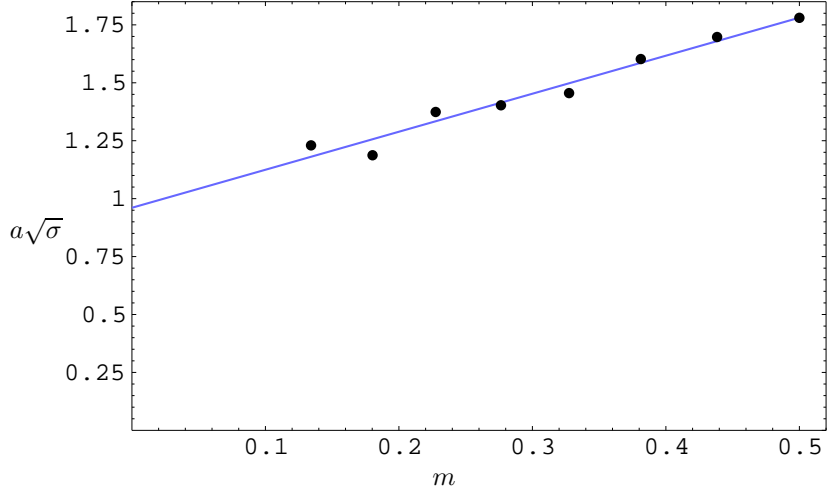 scaled 800}\\
&$m$
\end{tabular}
\caption{Variation of the lattice spacing $a$
vs.\ $m$ along the scaling trajectory of Table~\ref{tabtraj1}. 
Also shown is a fit to $0.96+1.64 m$.
\label{spacefig}}
\end{figure}

In Figures \ref{disp1} and \ref{disp2}, we plot the dispersion relation
for the $0^{++}$ and $0^{--}$ for a point on the scaling trajectory. 
This is a consistency check, to show whether optimizing the
slope at $|{\bf P}| = 0$ leads to covariant behaviour throughout the
Brillouin zone. 
Behavior in the figures is generic:  the dispersion of the $0^{++}$ is
extremely straight throughout the scaling trajectory and the 
dispersion of the $0^{--}$, along with the other excited states, tend
to follow a lattice dispersion $\propto \sqrt{1-\cos(a |{\bf P}|)}$
more closely.


\begin{table}
\begin{center}
\begin{tabular}{c|cccccc}
$\displaystyle {\left|{\cal J}\right|}^{{\cal P}_1 \cal C}$& 
\bBoxedEPSF{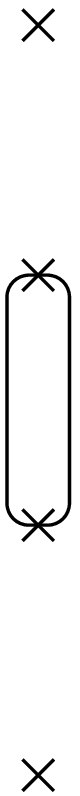 scaled 400}&
\bBoxedEPSF{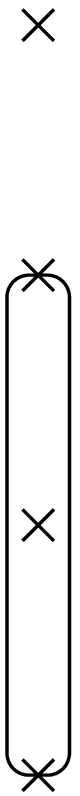 scaled 400}&
\bBoxedEPSF{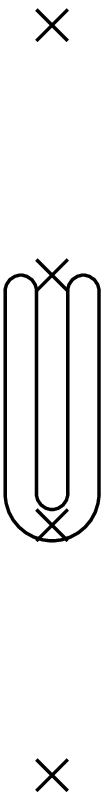 scaled 400}&
\bBoxedEPSF{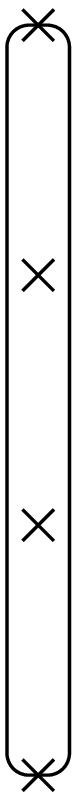 scaled 400}&
\bBoxedEPSF{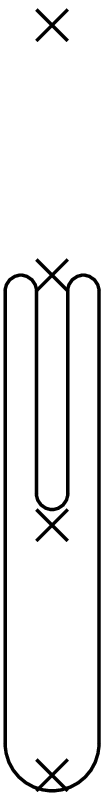 scaled 400}&
\bBoxedEPSF{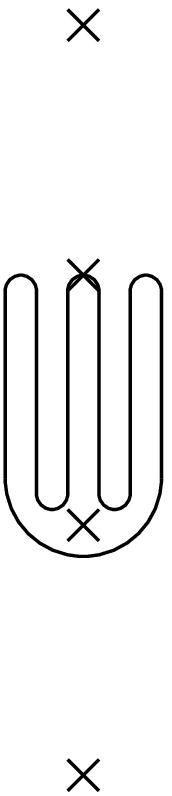 scaled 400}\\[10pt]
\hline
$0^{++}$ & 0.007   & 0.184 & 0.676 & 0.002 & 0.022 & 0.108 \\
$0^{++}_{*}$ & 0  & 0.716 & 0.189 & 0.009 & 0.027 & 0.059 \\
$0^{--}$ & 0.889  & 0.089 & 0.018 & 0.001 & 0.002 & 0.002     \\
$0^{--}_{*}$ & 0.069  & 0.68 & 0.118 & 0.023 & 0.053 & 0.056 \\
$2^{++}$ & 0.76   & 0.061 &0.14 & 0.001 & 0.007 & 0.03      \\
$2^{-+}$ & 0     & 0.871  & 0.002 & 0.005 & 0.061 & 0.061     \\
\end{tabular} 
\end{center}
\caption{
The transverse structure of each glueball for
$K=8$, $p \le 6$, at $m=0.134$, $\newl_1 = -0.153$,
$\newl_2 = -0.105$, $\newl_3  = 100$.
Each column shows the probability for a glueball to
contain a loop of given transverse shape. Information on the
relative phases of different shapes 
can similarly be obtained from the wavefunctions.
\label{looper}}
\end{table}

\begin{figure}
\centering
\BoxedEPSF{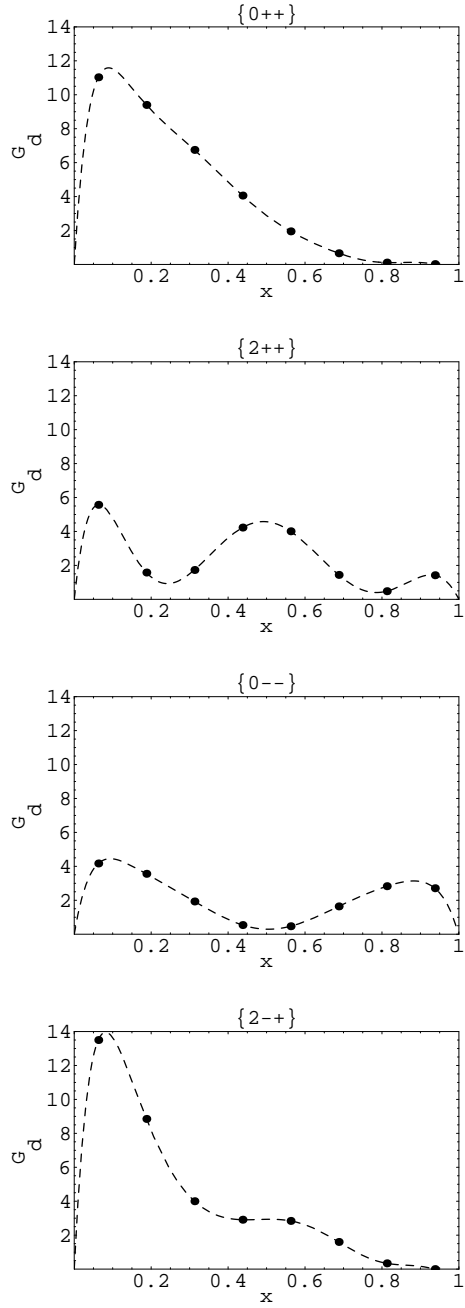 scaled 800}
\caption{
The structure functions $G_d(x)$ for 
$K=8$, $p \le 6$, at $m=0.134$, $\newl_1 = -0.153$,
$\newl_2 = -0.105$, $\newl_3  = 100$.
This exhibits the longitudinal structure of each glueball.
\label{structfig}}
\end{figure}

We can also deduce the transverse 
lattice spacing $a$ in physical units $\sqrt{\sigma}$, as one moves
along the trajectory of Table~\ref{tabtraj1} (see Fig.~\ref{spacefig}). 
This confirms that the
lattice spacing gradually decreases with $m$, and is always physically
quite large in the colour-dielectric regime $m^2 > 0$. Further information
can be gleaned from the glueball wavefunctions, which of course come
for free in a Hamiltonian calculation. In Table~\ref{looper} and  
Fig.~\ref{structfig}  we exhibit the
transverse and longitudinal structure of the glueballs at a representative
point in coupling space. We see a characteristic rise of the
$G_{d}(x)$-distribution at small $x$, due to higher Fock states. However, one
can show analytically that such distributions must vanish at $x=0$ for
two-dimensional gauge theories \cite{me} (without Yukawa interactions), 
and therefore
our colour dielectric regime $m^2 > 0$ in particular.

The coupling $\newl_3$, according to the test for Lorentz covariance, is
much larger than other couplings, which might appear unnatural at first
sight. However the corresponding operator is also distinguished by the
fact that at large $N$ it couples only to link--anti-link Fock states
(it `pinches' the flux tube swept out light-front time $x^+$). More 
precisely, if the link--anti-link wavefunction is $f(k^+,P^+ - k^+)$, the
$\newl_3$ operator has expectation value
\be
\langle (\Tr\left\{ MM^{\da} \right\})^2 \rangle \propto
 \left( \int_{0}^{P^{+}} dk^+ 
{f(k^+, P^+ - k^+) \over \sqrt{k^+ (P^{+} - k^+)}}
\right)^2 \label{pinch}
\eq
A complete set of link--anti-link wavefunctions $f$ resulting from the 
longitudinal Coulomb interaction can be labelled by the number of zeroes
of $f$ \cite{bard1}. 
The energy increases with the number of zeroes, and one might
naively expect the $0^{++}$ glueball to therefore be predominantly 
composed of a constant two-link wavefunction. However, this wavefunction
and only this wavefunction makes a significant contribution to  (\ref{pinch}).
A large $\newl_3$ therefore removes it from all low-lying glueballs, as is
evidenced from Table~\ref{looper} and fig.~\ref{structfig}. 
An intuitive explanation for this
follows from the fact that a wavefunction constant in $k^+$ implies
tight binding in the longitudinal direction. On the other hand, the
link--anti-link state has width $a$ in the transverse direction. From
Fig.~\ref{spacefig}
 we know this to be physically quite large (of order the glueball
width!). Therefore it is perhaps not surprising that the Lorentz
covariance conditions reject such an asymmetrical Fock wavefunction in
low-lying glueballs. If this conclusion  carries over to $3+1$
dimensions, it will result in a quite different picture from the one
suggested in Ref.~\cite{bard2} (which did not include a $\newl_3$ coupling).

The absence of the link--anti-link state means that the $0^{++}$ groundstate
glueball is predominantly formed from a linear combination of the 4-link
operators ${\cal O}_1 = \Tr\left\{ M^{\da}MM^{\da}M \right\}$ and
${\cal O}_{2} = \Tr\left\{ MMM^{\da}M^{\da} 
\right\}$. In fact, we find that the
$(0^{++},0^{++}_{*})$ system is to a good approximation described by
the symmetric and antisymmetric combinations 
$(\alpha {\cal O}_{1} + \beta {\cal O}_{2}, \alpha {\cal O}_{1} - 
\beta {\cal O}_{2})$. The operator ${\cal O}_{2}$ couples to $\newl_2 < 0$ in
the Hamiltonian and is the most significant hopping term on the
transverse lattice. As a result, we find the symmetric combination
lower in energy and with correct relativistic dispersion $c_{T}^{2} > 0$.
It then follows unavoidably that the higher-energy
anti-symmetric combination will have
an incorrect dispersion $c_{T}^{2} < 0$. Since we tentatively
identified the anti-symmetric combination with the $0^{++}_{*}$,
this explains why we always have a problem with this state. In fact,
it is almost purely lattice artifact at present. It is likely that the
effective potential must be extended to 6-link operators before
this problem can be resolved.

\subsection{Improving Higher Glueballs}

\begin{figure}
\centering
\BoxedEPSF{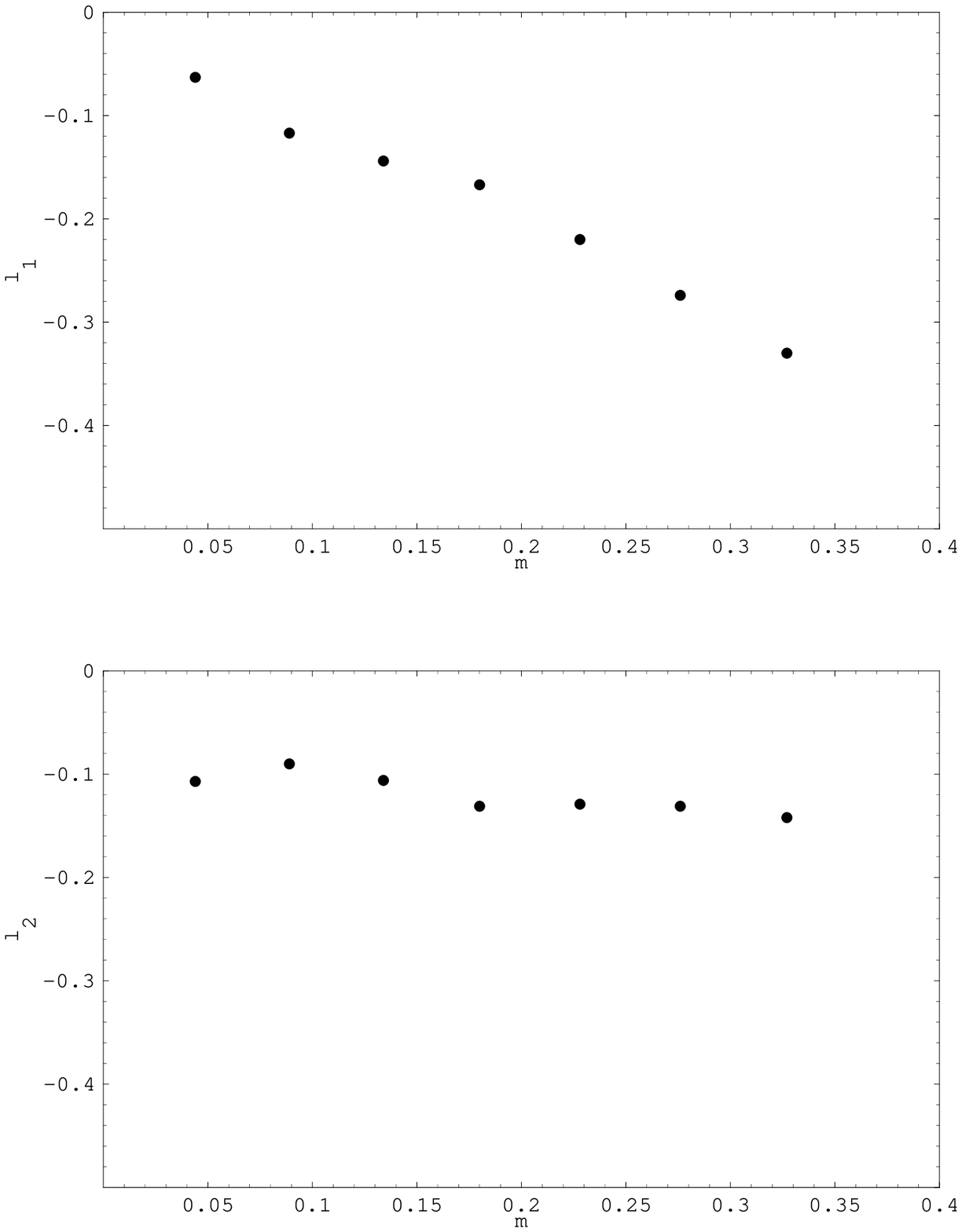 scaled 850}
\caption{The trajectory which minimises the $\chi^2$ test of Lorentz
invariance involving only glueball dispersion formulae. 
\label{trajn}}
\end{figure}
\begin{figure}
\centering
\BoxedEPSF{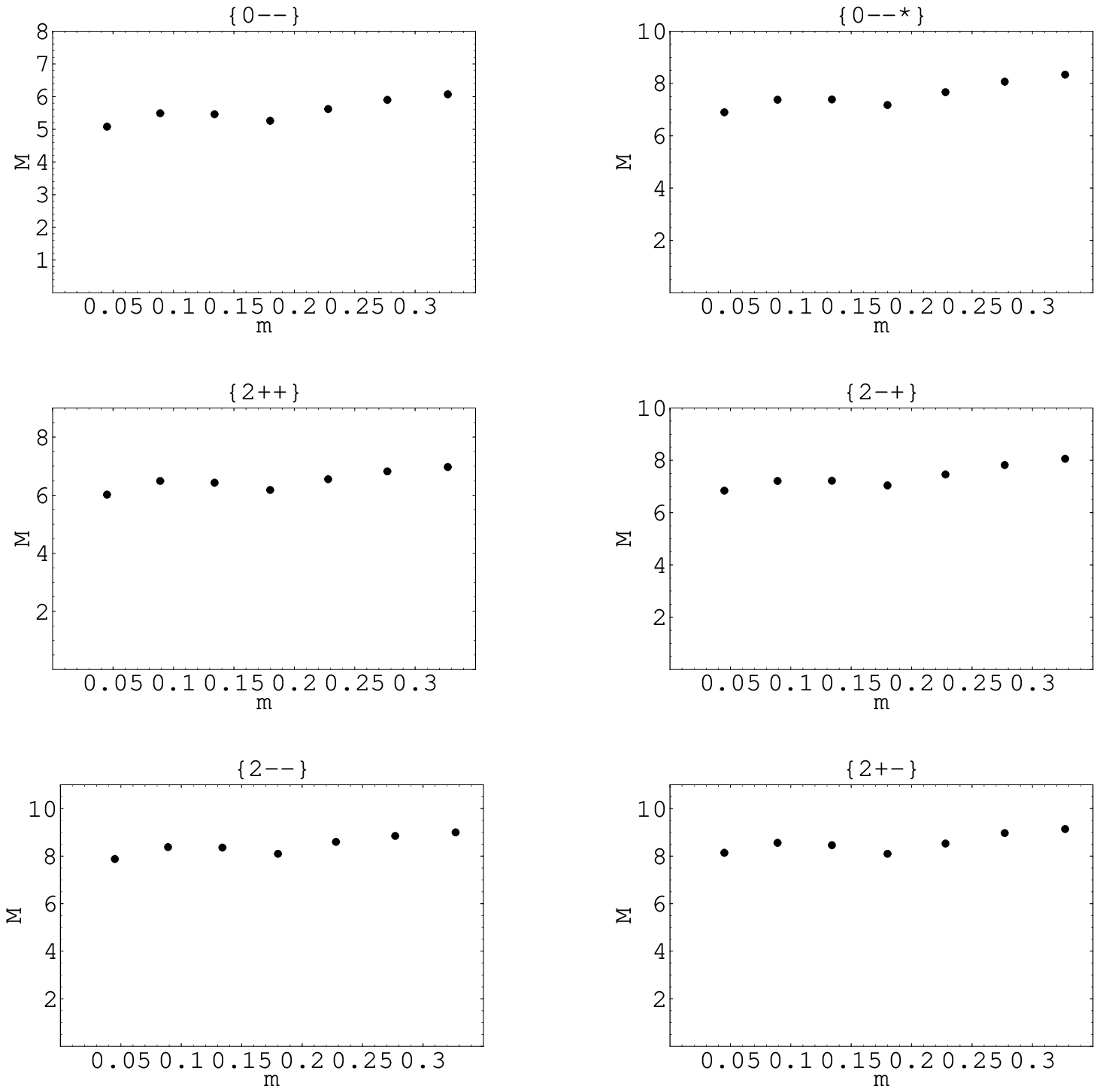 scaled 850}
\caption{Scaling of the higher glueball masses ${\cal M}$ in units
of $\sqrt{\sigma}$, 
as $m$ is varied along
the scaling trajectory of Fig.~\ref{trajn}. 
This gives an estimate of systematic errors for
each glueball from truncation of $V$.
\label{scaling}}
\end{figure}
We have found in the previous section that, determining the glueball masses
in terms of a scale set by the rotationally invariant string tension(s), 
we can obtain a reasonable
groundstate $0^{++}$, but cannot obtain Lorentz covariant higher glueballs  
in general. We attribute this incompatibility to our truncation of
the effective potential. It is natural to ask if one could find
 a coupling trajectory which improves the Lorentz covariance of higher
glueballs at the expense of rotational invariance of the string tension.
In particular, the longitudinal string tension is by far the most
difficult measurement we make, and a potential source of systematic
errors. To this end, we searched for such a trajectory, using the ELMC
value of $M_{0^{++}} / \sqrt{\sigma}$ as a phenomenological input to set
the scale. $\sigma$ was identified with $\sigma_T$ in this case. 
Although no longer quite a first principles calculation, the results are
sufficiently impressive, we feel, to display here.

\begin{figure}
\centering
\BoxedEPSF{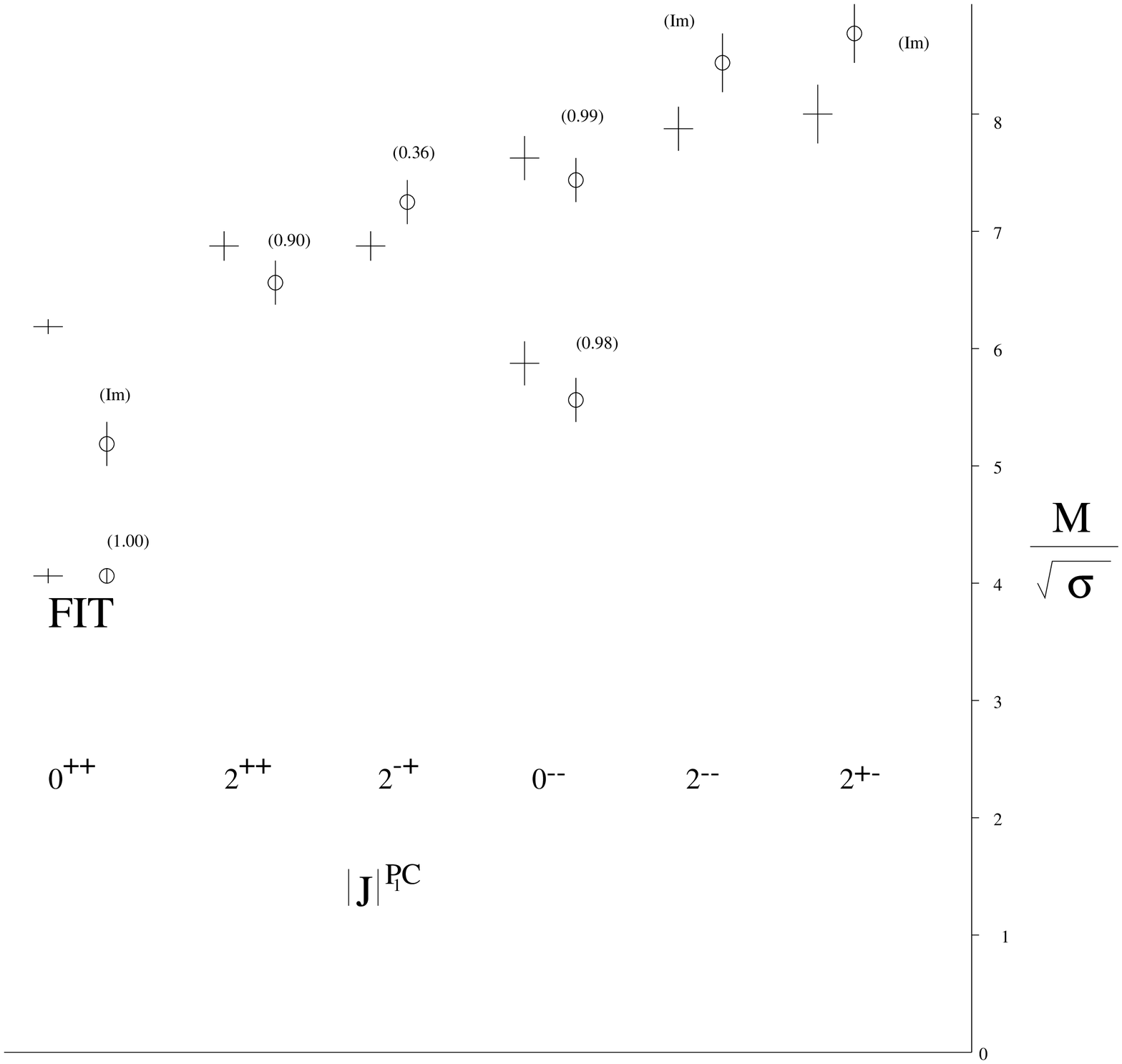 scaled 600}
\caption{The low-lying $2+1$-dimensional large $N$
glueball spectrum obtained on the transverse lattice
({\large o}). Also shown for comparison, the 
ELMC simulations of Teper (+), extrapolated to $N=\infty$ by a 
fit to $A + B/N^2$, with statistical
errors \cite{teper}. The value of $M_{0^{++}}/ \sqrt{\sigma}$ was fit to
the ELMC result, to set the scale. The other masses result from requiring
rotationally invariant dispersion, quantified by the speed
of light ratio $c_{T}/c_{L}$ deduced from each glueball (in brackets).
The estimated error from extrapolation in $K$ and $p$ is shown
but an estimate of systematic error from truncation
of $V$ is not shown (see Fig.~\ref{scaling}). 
\label{specfig}}
\end{figure}

Using a $\chi^2$ criterion which includes only the $c_{T}/c_{L}$ ratio
of the $0^{++}$ (whose mass is now fixed) and the $0^{--}$, we found the
trajectory displayed in Fig.~\ref{trajn}.
In fact, along this trajectory  not only 
are these
two glueballs rotationally invariant, but a number of others also have
a good $c_{T}/c_{L}$ ratio. The 
trajectory is close to the previous one, but rather more stable. The scaling
behaviour of glueball masses, shown in Fig.~\ref{scaling}, is also  more
stable. At the point of minimum $\chi^2$ on this trajectory, we find the
spectrum displayed in Fig.~\ref{specfig}. 
In particular, it is worth noting that
those glueballs whose $c_{T}/c_{L}$ ratio is close to one have predicted
masses in agreement with the ELMC result.  
This encourages us that, by adding further couplings to the effective 
potential to improve the Lorentz covariance of higher glueballs (from
first principles), should bring their masses into line also.


\section{Conclusions.}

We have constructed an effective transverse lattice 
light-front Hamiltonian of pure gauge theory, 
extending the discussion
of quantisation and the large-$N$ limit addressed in the earlier
works \cite{us1,bard1,bard2}, and outlined a first-principles
procedure for solving the eigenvalue problem. 
Not only does this formulation of gauge theory possess the
usual appealing features associated with light-front quantisation
--- trivial vacuum, boost invariance, parton dynamics --- 
but it also efficiently deals with 
the common problems found in this approach
--- zero modes, renormalisation, non-compactness/gauge invariance. 
It does this at the price
of introducing an effective potential which must be tuned
by demanding Lorentz covariance of eigenstates of the
light-front Hamiltonian at large transverse lattice spacing.
Further remarkable simplifications occur in the large
$N$ limit. Only a $1+1$ dimensional theory of connected Wilson loops
needs to be solved.  Equivalently, one solves a $1+1$ dimensional 
continuum gauge theory coupled to complex adjoint matter fields 
which form freely propagating colour strings.

We successfully tested the ideas by performing comprehensive calculations
in $2+1$ dimensions at large $N$, using an improvement scheme to
extrapolate to the longitudinal continuum limit.
We found agreement with conventional Euclidean lattice
simulations \cite{teper}, 
in those observables for which Lorentz covariance was attained.
Furthermore,
we could go beyond the path integral approach by obtaining the
frame-invariant
glueball wavefunctions.
Given that the quantisation, regulators, elementary fields, and gauge
fixing are different from those of Ref.~\cite{teper}, the agreement is
quite impressive. It non-perturbatively 
validates both the lattice colour-dielectric formulation 
and the light-front Hamiltonian quantisation of 
non-Abelian gauge theory at the quantitative level. 
This is therefore an important step in the application of light-front
quantisation to hadronic physics in general.

The obvious extension of large-$N$ 
pure glue calculations to $3+1$ dimensions has 
already begun, and initial results are encouraging. For couplings
in the light-front Hamiltonian (\ref{redham}) similar to those found for 
the $2+1$ problem, we find $0^{++}, 2^{++},$ and $1^{+-}$ multiplets
as the lightest
glueballs (only some of the components of these multiplets were found in
the
original work \cite{bard2}). A search of the full parameter space
for a scaling trajectory will yield estimates of the masses and
wavefunctions.
Given that $1/N^2$ corrections are expected to be smaller than other 
systematic errors, 
the masses can be compared with various experimental glueball candidates,
providing a useful alternative to the conventional ELMC estimates.

\vspace{10mm}

\noindent
{\bf Acknowledgements:} SD thanks 
Prof.\ F. Lenz (Erlangen) and Mrs.\ van de Sande
for hospitality during part of the work, and M. Teper for many helpful
interactions.
The work of SD was supported  by a Particle Physics and 
Astronomy Research Council (U.K.) Advanced Fellowship and a CERN fellowship.
\vspace{5mm}


\newpage
\begin{appendix}

\section{$A_{-}$ Zero Modes}
\label{appendixa}

\begin{figure}
\centering
\BoxedEPSF{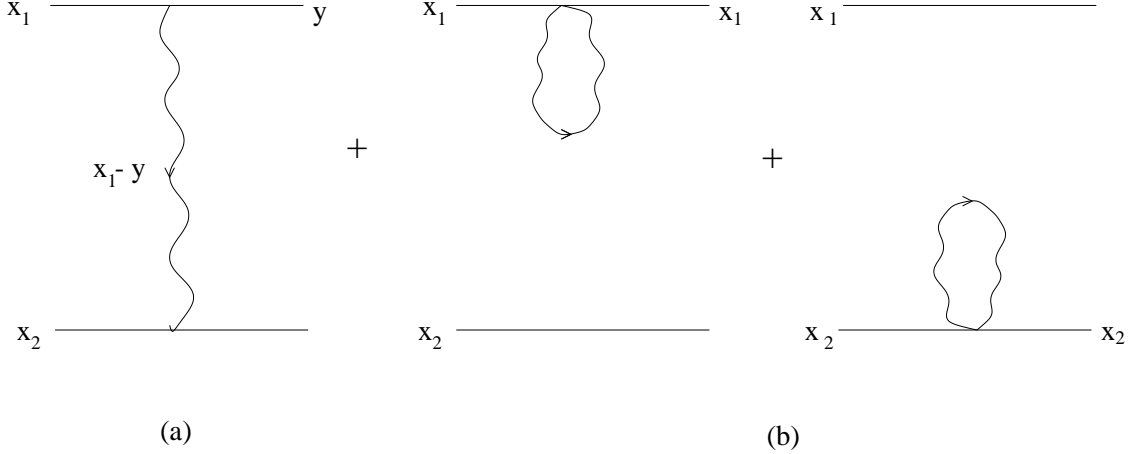 scaled 400}
\caption{(a) Coulomb exchange graph; (b) self-energy graphs (contraction
of (a).
\label{fign}}
\end{figure}

In this appendix, we argue that the only effect of the $A_{-}$ zero modes
$c_i$ in the longitudinal Coulomb interaction (\ref{cou})
is to finitely renormalise the link field mass $\mu^2$.
The two processes
contributing to  Coulomb exchange between link-variables $M_r({\bf x})$
in light-front Hamiltonian bound state equations are
shown diagrammatically in Fig.~\ref{fign} (see for example Refs.~\cite{model}
for further discussion). Fig.~\ref{fign}(a) represents the
instantaneous exchange of the $A_{+}$ non-zero mode, while Fig.~\ref{fign}(b)
is the self-energy contribution arising from normal-ordering the former.
Neglecting $c_i$, Fig.~\ref{fign}(b) acts to regulate the singular behaviour
of Fig.~\ref{fign}(a) at zero momentum transfer,
by an equivalent  principle value prescription \cite{bard1}.
Let us denote the relevant Fock wavefunction 
of the two link variables 
by $ f(\cdots, x_1, x_2, \cdots)$ (or $f(x_1,x_2)$ for short),
for neighboring links in the colour Trace of a string state
(e.g.\ref{typ}), carrying momenta
$x_1 P^+$ and $x_2 P^+$ in the boundstate $|\Psi (P^+) \rangle$ of total
momentum $P^+$. As explained in section~\ref{focsec}, only neighboring
links in the colour Trace have Coulomb exchange interactions in the
large-$N$ limit. We can neglect labels for the link variables other 
than longitudinal momentum. 
Our basic assumption in the following is that
in order to estimate the effect of zero modes $c_i$, which are constrained
variables at large $N$ whose distribution $\phi(c_1,\ldots,c_N)$
is in principle known,
we insert a c-number $C/{\cal L}$ 
wherever a quantity like $c_i / {\cal L}$ would 
appear in the
exact expression at finite ${\cal L}$. In particular,
$C$'s appearing in any expression 
should not be identified numerically
with one another, but each represent something $O(1)$ 
in the continuum limit ${\cal L} \to \infty$.

The contribution of the Coulomb interaction to the
matrix element of the light-front Schr{\"o}dinger operator
$\langle {\rm Fock} | 2P^+ P^- | \Psi \rangle$ is of the form\footnote{Here
we neglect to show mass renormalisations which are known already from the 
naive analysis \cite{bard2}.}
\begin{eqnarray}
 I(x_1,x_2) &=& {\cal L} \sum_{m=1}^{n_1 + n_2} 
{n_1 + m + C \over \sqrt{(n_1 + C)(m+C)}} {1 \over
(n_1 - m + C)^2} 
\nonumber\\ &&\hspace{0.1in}\cdot
 \left\{ {f(n_1, n_2) (n_1 + m + C) \over \sqrt{(n_1 + C)
(m + C)}} - {f(m, n_1 + n_2 -m) (2n_2 + n_1 -m+C) \over \sqrt{ (n_2 + C)
(n_1 + n_2 -m +C)}} \right\}\; .  \label{can}
\end{eqnarray}
Here $x_1 = n_1/K$, $x_2 = n_2 / K$, and $n_1,n_2,m\neq n_1$ are
non-zero integers less than integer $K =  {\cal L} P^+ / 2 \pi$.
(If we could set $C=0$ and take the limit ${\cal L} \to \infty$, 
we would recover the naive kernel of Ref.~\cite{bard1}, which 
employed the naive light-front gauge $A_{-} = 0$.)
For a finite free kinetic energy in the colour-dielectric regime,
$n_1$ and $n_2$ must be both $O(\cal L)$.
We Taylor expand (\ref{can}) in $C/{\cal L}$ to obtain
\begin{eqnarray}
I(x_1,x_2) & = & {\cal L} \sum_{m'= -n_1}^{n_2} 
{2n_1 + m'  \over \sqrt{n_1 (n_1 + m')}} {1 \over (C-m')^2}
\nonumber \\ & & \hspace{0.1in} \cdot
 \left\{ { f(n_1, n_2) \over \sqrt{n_1(n_1 + m')}}
 \left( 2n_1 + m' - C \left(1- {(2n_1+m')^2 \over n_1 (n_1 + m')} \right)
\right) \right. \nonumber \\   && \hspace{0.2in} 
\left. - {f(n_1 +m',n_2 - m')  \over \sqrt{n_2(n_2 - m')}}
 \left( 2n_2 - m' - C \left(1- {(2n_2 - m')^2 \over n_2 (n_2 - m')} \right)
\right) \right\}  
\nonumber\\ & & \hspace{0.1in}
+\, O\!\left(\frac{C}{{\cal L}}\right)\; . \label{taylor}
\end{eqnarray}
The higher corrections $O(C/ {\cal L})$ can be dropped in the
continuum limit. 
We consider first the term at $O(C)$ in the curly brackets of (\ref{taylor}).
As ${\cal L \to \infty}$ it can only affect the finite part of $I$
through the region of infinitesimal momentum transfer $m' = m - n_1 
\sim O(1)$. In this region it simplifies to
\begin{eqnarray}
{\cal L} f(n_1, n_2) \left( {C \over n_1} + {C \over n_2} \right)
 \sum_{m'= -n_1}^{n_2} 
{1 \over (C-m')^2}  
\end{eqnarray}
which, when folded with the appropriate distribution $\phi(c_1, \cdots, 
c_N)^2$, is equivalent to shifting the mass $\mu^2$ in boundstate
equations.
The key point is the 
absence of cross-terms like $1/\sqrt{x_1 x_2}$. To actually calculate
the mass shift would require $\phi$, which we have not tried
to determine in detail.
In the case of $SU(2)$ scalar matter, 
it should be possible numerically to check
whether the result of Ref.~\cite{pauli} can be reinterpreted as a mass
shift. Note that our argument does not lead to
any mass shift for fermionic matter. 

Turning to the term $O(C^0)$ in the curly brackets of (\ref{taylor}),
this reduces to the continuum kernel of Ref.~\cite{bard1} but for the
contributions in the region $m' \sim O(1)$. To discuss the latter, we
must specify more carefully the $C$'s. The correct expression for this
region is (c.f. (\ref{coul}))
\begin{eqnarray}
\int_{0}^{1} dc_i\, |\phi(c_1, \cdots , c_N)|^2 4 \,\Delta f(n_1, n_2)\,
{\cal L} \sum_{m'= -n_1}^{n_2} 
 { m'  \over (c_i - c_j -m')^2} + O(1/{\cal L})
 \label{newcou}
\end{eqnarray}
where we have now explicitly integrated over the  zero mode distribution
$\phi$, and used 
\be
f(n_1, n_2)  - f(n_1 +m',n_2 - m') = m' \, \Delta f(n_1, n_2)  + O(1/{\cal L}^2)
\eq
with $\Delta f(n_1, n_2) \sim O(1/{\cal L})$. At first sight it appears as
if the zero modes $c_i$ spoil the cancellation of (\ref{newcou}) between
positive and negative $m'$ which would occur in their absence. 
However, using the fact that $\phi$ has
definite symmetry under interchange of any two $c_i$'s, ones sees
that after integration over $c_i$'s, the cancellation does hold.
Another, more physical, way to see why this cancellation must occur, is
that a derivative term $df(x)/dx$ would result in the continuum bound state
equation if it did not. Such a term would violate $x^{D-1} \to -x^{D-1}$
symmetry in general.

\section{Nonzero winding and nonzero transverse momentum}
\label{nonzerowinding}

The analysis of nonzero transverse momentum in Section~\ref{lorentz}
can be extended to nonzero winding number ${\bf n}\neq 0$.
Thus one can also study the Lorentz boost properties of the
winding modes (we have not yet attempted systematic study of this).

Let us start with a $p$-link state that winds once around
a lattice with $\bf n$ transverse links.
A generic ${\bf P}=0$ state has the form
\begin{eqnarray}
   \left | \Psi(P^+,0)\right\rangle = 
    \frac{1}{\sqrt{V N^p}} \sum_{\bf y}
     \Tr\left\{ a_{\lambda_1}^\da(k_1,{\bf x}_1+{\bf y})  
        a_{\lambda_2}^\da(k_2,{\bf x}_2+{\bf y})
        \right. \nonumber \\ \left. 
        \cdots a_{\lambda_p}^\da(k_p,{\bf x}_p+{\bf y})\right\} 
        \left|0\right\rangle
 \label{afirst}
\end{eqnarray}
where $\bf y$ is summed over all $V=n^1 \cdots n^{D-2}$ sites of the lattice
and $\sum_i \widehat{\lambda}_i = {\bf n}$. By periodicity of the lattice 
we identify $a_\lambda^\da(k,{\bf x})$ with 
$a_\lambda^\da(k,{\bf x}+a {\bf n})$.
The transverse coordinates are given by the formula 
\be
{\bf x}_i = {\bf x}_{i-1} +a \,\widehat{\lambda}_{i-1} 
                   \;, \;\;\;\; 1<i\le p\; .
\eq
Note that the number of nonzero contractions 
$w=\langle  \Psi(P^+,0)| \Psi(P^+,0)\rangle$
must divide $n^r$ evenly, $n^r/w \in \integer$, for all $r$.
Now, we will boost this state (\ref{afirst}) to some finite transverse
momentum ${\bf P}$:
\begin{eqnarray}
   \left |  \Psi(P^+,{\bf P})\right\rangle 
   &=& {\rm e}^{i P^r M_{-r}/P^+} \left | \Psi(P^+,0)\right\rangle\nonumber\\  
        &=& \frac{1}{\sqrt{V N^p}} \sum_{\bf y}
        {\rm e}^{i \bf{P}\cdot{\bf y}} 
      \Tr\left\{ a_{\lambda_1}^\da(k_1,{\bf y})  
        a_{\lambda_2}^\da(k_2,{\bf x}_2+{\bf y})
        \cdots  a_{\lambda_p}^\da(k_p,{\bf x}_p+{\bf y})\right\} 
        \left|0\right\rangle
        \; . 
\end{eqnarray}
For nonzero winding, the phase condition introduced earlier (\ref{com})
has no particular advantage; instead we have chosen the convention 
${\bf x}_1={\bf \bar x}=0$.

Since we have a periodic lattice, ${\bf P}$ can only take
on discrete values
\be
        l^r \equiv \frac{a P^r n^r}{2 \pi} \in \integer \;\;\;\;
        \mbox{(no sum over $r$).}
\eq
We find the inner product
\begin{eqnarray}
        \langle \Psi(P^+,{\bf P})|\Psi(P^+,{\bf P})\rangle &=& 
               \langle \Psi(P^+,0)|\, {\rm e}^{-i P^r M_{-r}/P^+}
                 {\rm e}^{i P^r M_{-r}/P^+}
                 \,|\Psi(P^+,0)\rangle \nonumber \\
        &=&
                 \sum_{v=0}^{w-1} {\rm e}^{i a {\bf P}\cdot{\bf n} v/w} 
\nonumber \\
        &=& \left\{ \begin{array}{ll}
               w \; , & \sum_r l^r/w \in \integer \\
                0 \; , & \mbox{otherwise} \end{array} \right. \; .
\end{eqnarray}
%
Thus, as we change the transverse momentum, the number of states in
a basis can change.   Unlike the ${\bf n}=0$ case, 
the boost operator is no longer self-adjoint due
to the finite transverse volume of the system.

In practical calculations, we prefer to work with real numbers.
Thus, we introduce orientation reversals
\begin{eqnarray}
 \lefteqn{ {\cal O} \Tr\left\{ a_{\lambda_1}^\da(k_1,{\bf x}_1) 
         a_{\lambda_2}^\da(k_2,{\bf x}_2)
        \cdots  a_{\lambda_p}^\da(k_p,{\bf x}_p)\right\} \left|0\right\rangle
    }\nonumber \\
 &=& \Tr\left\{ a_{\lambda_p}^\da(k_p,{\bf y}_1)  
        a_{\lambda_{p-1}}^\da(k_{p-1},{\bf y}_2)
        \cdots  a_{\lambda_1}^\da(k_1,{\bf y}_p)\right\} 
                        \left|0\right\rangle \; .
\end{eqnarray}
We perform no cyclic permutations on the colour trace.
We find
\be
        {\bf y}_i = {\bf n}-{\bf x}_{p-i+1}
\eq
Consequently, 
${\cal O}\, {\rm e}^{i B}\,{\cal O}= {\rm e}^{-i B}$ 
where $B=P^r M_{-r}/P^+$.
It is useful to introduce the two states:
\begin{eqnarray}
      |A(P^+,{\bf P})\rangle & = & 
                \cos\left(B\right)(1+{\cal O})|\Psi(P^+,0)\rangle
        \nonumber \\
      |B(P^+,{\bf P})\rangle & = & 
                \sin\left(B\right)(1-{\cal O})|\Psi(P^+,0)\rangle \; .
\end{eqnarray}
For states without symmetry under orientation reversal
$\langle \Psi(P^+,0)|{\cal O}|\Psi(P^+,0)\rangle =0$, 
we find the inner products
\begin{eqnarray}
        \langle A(P^+,{\bf P}) | A(P^+,{\bf P})\rangle =
        \langle B(P^+,{\bf P}) | B(P^+,{\bf P})\rangle &=& 
                \left\{ \begin{array}{ll}
               w \; , &  \sum_r l^r/w \in \integer \\
                0 \; , & \mbox{otherwise} \end{array} \right.  \\
        \langle A(P^+,{\bf P}) | B(P^+,{\bf P})\rangle = 
        \langle B(P^+,{\bf P}) | A(P^+,{\bf P})\rangle &=& 0 
\end{eqnarray}
For states which are symmetric under orientation reversal 
$\langle \Psi(P^+,0)|{\cal O}|\Psi(P^+,0)\rangle\neq 0$, 
$| A(P^+,{\bf P})\rangle $ and $| B(P^+,{\bf P})\rangle $
are equivalent up to a phase factor.  In 
this case, we include only the $| A(P^+,{\bf P})\rangle $ state in 
the basis.  
%
Occasionally, $\langle A(P^+,{\bf P}) | A(P^+,{\bf P})\rangle=0$
in which case we include $|B(P^+,{\bf P})\rangle$ in the basis instead.


\section{Improved matrix elements for DLCQ}
\label{appendixc}

In this section, we will describe some refinements to the basic 
technique introduced in Appendix~C of \firstpaper{}.
Our method is the following:  when calculating the matrix elements
of the Hamiltonian in a many-particle calculation, we will use 
ordinary DLCQ commutation relations 
to calculate contractions associated with spectator particles.
To calculate the interaction itself, we take the DLCQ momenta of
the interacting particles, project onto a smooth wavefunction basis
(continuous longitudinal momentum), and calculate matrix elements
of the interaction in this smooth wavefunction basis.  
Our modification of DLCQ essentially eliminates the 
$1/\sqrt{K}$ and $1/K^{2 \beta}$ errors associated with small momenta.
The remaining errors include a (small) $1/K$ error from the
instantaneous interaction $1/(k^+-p^+)^2$, $1/K^2$ errors from
DLCQ itself, and a $1/K$ error from production of
particles at small momentum.
In the following, we discuss improvements to 
the $(\mbox{2 particles}) \to (\mbox{2 particles})$ interactions,
the $(\mbox{1 particle}) \to (\mbox{3 particles})$ interactions,
and the heavy source interactions. The `particles' in the present
context are the link fields of course.

\subsection{$(\mbox{2 particles}) \to (\mbox{2 particles})$ interactions}

Consider a typical 4-point interaction
%
%
\begin{equation}
V\!\left(\frac{k_1}{K},\frac{k_3}{K}\right)=
\begin{array}{r}k_1\\[25pt]k_2 \end{array}
\BoxedEPSF{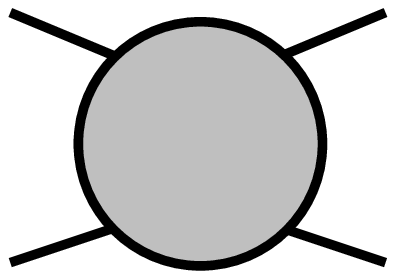 scaled 500}
\begin{array}{l}k_3\\[25pt]k_4\end{array}
     \label{vtwo}
\end{equation}
with DLCQ momenta $k_i \in \{1/2,3/2,\ldots\}$ and 
$K = k_1 +k_2 = k_3 + k_4$.
In particular, the mass term is treated as if it were a 4-particle
interaction:
\be
  V(x,y) = \frac{\mu^2}{2}\left(\frac{1}{x}+\frac{1}{1-x}\right) 
                \delta(x-y)\; . \label{imass}
\eq

We start by calculating eigenfunctions of the two-particle 
bound state equation~\cite{bard1}.
\begin{eqnarray}
  \lefteqn{{\cal M}^2 f(x,1-x)} \nonumber \\
  &=& \int_0^1 dy\,H(x,y) \, f(y,1-y)\nonumber \\
  &=& \frac{\mu^2\, f(x,1-x)}{x (1-x)}
    - \frac{G^2 N}{4 \pi} \int_0^1 dy\, \frac{ f(y,1-y)}{\sqrt{x (1-x) y
   (1-y)}}\left(\frac{(x+y)(2-x-y)}{(x-y)^2} +1\right)
  \label{twoe}
\end{eqnarray}
where $\lambda_1 +\lambda_2/2+\lambda_3 = - G^2 N a^{D-2}$. 
Then we write the incoming and outgoing DLCQ states 
as linear combinations of the first $K$ eigenfunctions and 
evaluate matrix elements of the various 
interactions $V(x,y)$ in the eigenfunction basis. 

We introduce a basis of even and odd wavefunctions
(labeled by subscripts $s$ and $s^\prime$):
\begin{equation}
  \phi_s\!(x) = (x(1-x))^\beta \left\{\begin{array}{cc}
                \cos\left(\pi s\left(x-\frac{1}{2}\right)\right)\;, 
                & s\in\{0,2,\ldots,2 S\}\\
                \sin\left(\pi s\left(x-\frac{1}{2}\right)\right)\;, 
                & s\in\{1,3,\ldots,2 S-1\}\
        \end{array}
               \right. \; ,
\end{equation}
with $\beta$ is given by $\mu^2=G^2 N \beta  \tan(\pi \beta)$, $0<\beta<1/2$.
We define, for both even and odd wavefuntions, 
matrix elements of the various interaction terms  
$V_{s,s^\prime} = \int dx\, dy \, \phi_s (x)\, V\left(x,y\right)\, 
\phi_{s^\prime} (y)$, 
the inner product matrix 
$E_{s,s^\prime} = \int dx\, dy \, \phi_s (x)\, \phi_{s^\prime} (y)$, and
the two-particle Hamiltonian
$H_{s,s^\prime} = \int dx\, dy \, \phi_s (x)\, H\left(x,y\right)\, 
                        \phi_{s^\prime} (y)$.  
In addition, we define 
the matrix $T_{i,s}$ which maps the 
wavefunction basis onto the DLCQ basis 
\begin{equation}
  T_{i,s} =\left(\frac{k_1 k_2}{K^2}\right)^\beta
        \left\{\begin{array}{c}
                \cos\left(\pi s\left(\frac{k_1-k2}{2 K}\right)\right) \\
                \sin\left(\pi s\left(\frac{k_1-k2}{2 K}\right)\right) 
             \end{array}\right.
\end{equation}
where the subscript $i \in \{1,2,\ldots\,K\}$
denotes a two particle DLCQ basis state with momenta 
$k_1=i-1/2$, $k_2=K-k_1$.

This choice of basis wavefunctions has several advantages
over our previous choice.  First, this basis is numerically
`well-conditioned' while the previous choice was numerically
unstable.  Second, the highly excited eigenfunctions of (\ref{twoe})
are themselves nearly sines and cosines.
We now generate the improved matrix elements at the beginning
of a calculation, store them in memory, and use them as needed.

One calculates the various integrals using the formula
\be
        \int_0^1 dx \, x^\mu (1-x)^\mu\, \cos\left(2 a\left(x-1/2
                \right)\right) = \sqrt{\pi}\,\Gamma(\mu+1)\, f_{\mu+1/2}(2 a)
\eq
where $f_\mu(x)$ is defined in terms of a Bessel function:
\begin{eqnarray}
        f_\mu(z) &= & J_\mu(z/2)/z^\mu \\
        \frac{d}{dx} f_\mu(x)&=& -\frac{x}{2} f_{\mu+1}(x)\\
        4 \mu f_\mu(x)&=&f_{\mu-1}(x)+ x^2 f_{\mu+1}(x)\\
        \lim_{x\to 0} f_{\mu}(x) &=& \frac{1}{\Gamma(\mu+1) 2^{2 \mu}} \; .
\end{eqnarray}
Thus, for example,
\be
        E_{s,s^\prime} = \frac{\sqrt{\pi}}{2}\,\Gamma(2 \beta+1)
                \left[ f_{2 \beta+\frac{1}{2}}\left(\pi (s-s^\prime)\right) 
                \pm  f_{2 \beta+\frac{1}{2}}\left(\pi (s+s^\prime)\right)
                \right] \; .
\eq
The greatest difficulty comes from the instantaneous interaction
\be
  V_{s,s^\prime} = -\frac{G^2 N}{8 \pi} \int dx\, dy\,
                \frac{(x+y)(2-x-y) \,\phi_s(x)\,\phi_{s^\prime}(y)}{
                \sqrt{x y (1-x) (1-y)} (x-y)^2} \; . \label{vv1}
\eq
We cannot solve this integral analytically, but we can subtract the
pole part at $x=y$ and solve the remaining finite integral numerically.
Thus, we add and subtract the quantity
\begin{eqnarray}
& &-\frac{G^2 N}{8 \pi} \int dx\, dy\,
                \frac{(x+y)(2-x-y) \left(x (1-x) y (1-y)\right)^{\beta-1/2}}{
                (x-y)^2} 
        \nonumber\\ & & \hspace{0.5in}
          \cdot \left[\frac{1}{2}
                \cos\left[\frac{\pi}{2}(s-s^\prime)(x+y-1)\right]
                \pm\frac{1}{2}
                \cos\left[\frac{\pi}{2}(s+s^\prime)(x+y-1)\right]
                \right] \label{vv2}
\end{eqnarray}
which, using the identity
\be
  \frac{1}{(x-y)^2}= -\lim_{\epsilon\to 0} \int_0^\infty dz\,
        z \cos\!\left(z (x-y)\right)\,{\rm e}^{-\epsilon z}\; ,
\eq
equals
\begin{eqnarray}
 & & G^2 N \frac{\Gamma(\beta+1/2)}{16}\left\{ 
                f_\beta\!\left(v^-\right) \left[
                (1+4\beta)\, f_\beta\!\left(v^-\right)
                -2 \left(v^-\right)^2 
                  f_{\beta+1}\!\left(v^-\right)\right]
        \right. \nonumber \\ & & \hspace{0.75in} \left.
        \pm f_\beta\!\left(v^+\right) \left[
                (1+4\beta) \, f_\beta\!\left(v^+\right)
                -2 \left(v^+\right)^2 
                  f_{\beta+1}\!\left(v^+\right)\right]
                \right\}
\end{eqnarray}
where $v^\pm = \pi (s\pm s^\prime)/2$.  The numerical integration
of (\ref{vv1}) minus (\ref{vv2}) is performed using the
Gau{\ss}-Jacobi method~\cite{numerical} which correctly handles the endpoint 
behavior of the integrand and converges exponentially fast.

We solve the generalised eigenvalue problem
$H \chi_t = h_t E \chi_t$ where the eigenvectors
$\chi_t$ have normalisation
$\chi_{t^\prime}^\da E \chi_t=\delta_{t,t^\prime}$.
Using the lowest $K$ eigenvectors, 
we define an eigenstate basis
(denoted by subscripts $t$ and $t^\prime$),
\begin{equation}
T_{i,t}=T_{i,s} \left(\chi_t\right)_s \;  .
\end{equation}
To make the transformation between the DLCQ and 
eigenfunction bases orthonormal, we perform a standard
QR-factorisation~\cite{numerical}, $T_{i,t}=Q_{i,t^\prime} R_{t^\prime,t}$, 
and discard the triangular
matrix $R$ (diagonal elements of $R$ must be positive).
Finally, we define the improved matrix elements by
\be
    V_{i,j} = Q_{i,t} \left(\chi_t\right)^\da_s
            V_{s,s^\prime} \left(\chi_{t^\prime}\right)_{s^\prime} 
                Q^T_{t^\prime,j}
          \; .
\eq

\subsection{$(\mbox{1 particle}) \to (\mbox{3 particles})$ interactions}

\begin{figure}
\centering
\BoxedEPSF{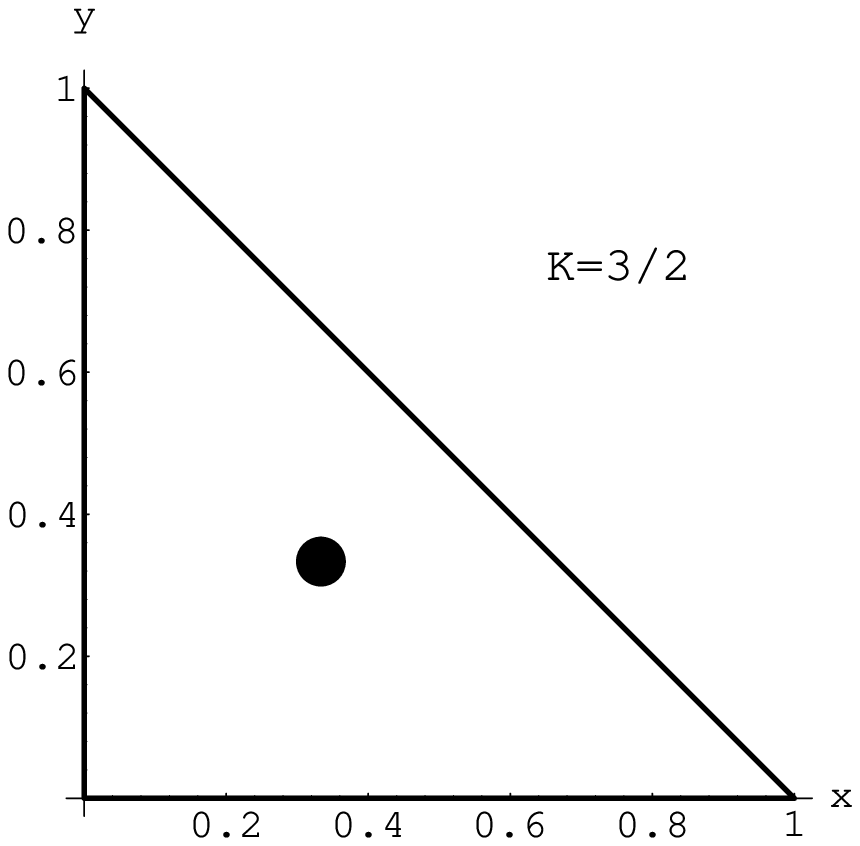 scaled 550}\hfill
\BoxedEPSF{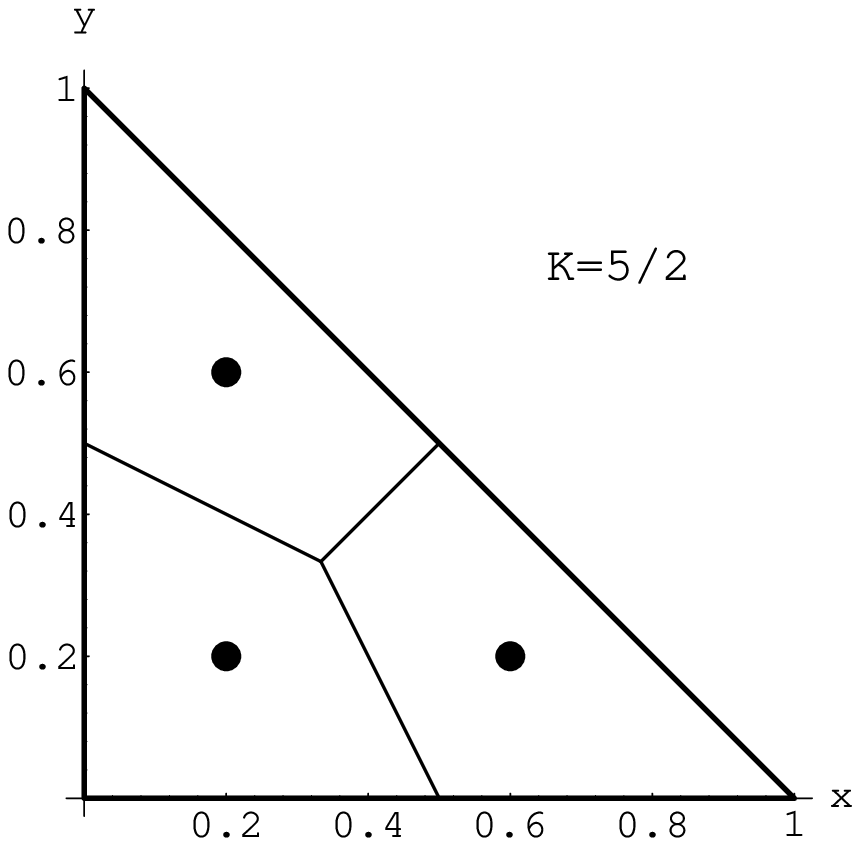 scaled 550}\hfill
\BoxedEPSF{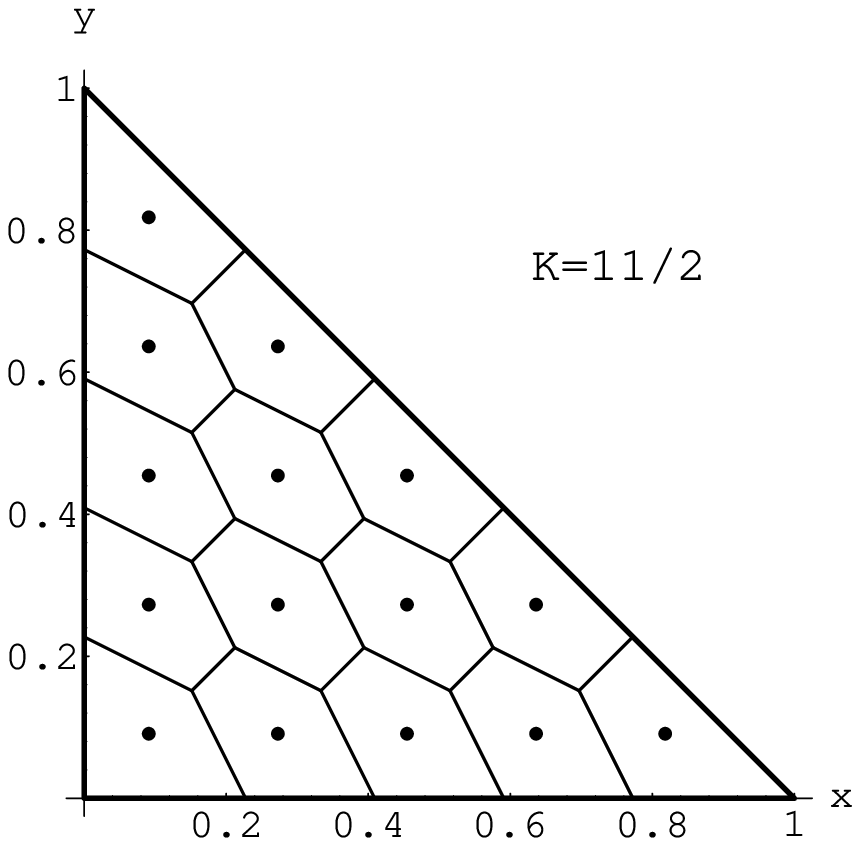 scaled 550}
\caption{Region $R_{k_1,k_2,k_3}$ for $K=3/2$, $K=5/2$, and $K=11/2$.
  The dots represent the points $(k_1/K,k_2/K)$ and $R_{k_1,k_2,k_3}$
    is the surrounding polygonal region.
   \label{triangles}}
\end{figure}

{}From the general bound state equations of two-dimensional gauge theories
with pair production, one can show that as
the momenta of any {\em two} particles go to zero
the wavefunction becomes constant \cite{me}. 
Thus we define a basis of wavefunctions
\be
    \phi_{k_1,k_2,k_3}(x,y)=  C_{k_1,k_2,k_3} \left\{\begin{array}{cl}
            \displaystyle \frac{\left(x y (1-x-y)\right)^\beta}{
                \left((x+y)(1-x) (1-y) \right)^{2 \beta}}
           \; , & (x,y)\in R_{k_1,k_2,k_3} \\
           \displaystyle 0 \; , & \mbox{otherwise}
         \end{array} \right. \; .
\eq
with normalisation
\be
    1= \int_0^1 dx \int_0^{1-x} dy \, {\phi_{k_1,k_2,k_3}(x,y)}^2
\eq
for three outgoing particles with longitudinal momentum fractions
$x$, $y$, and $1-x-y$ in the continuous wavefunction basis
and $k_i \in \{1/2,3/2,\ldots\}$, $K = k_1 +k_2 + k_3$,
in the DLCQ basis.
We define the polygonal region $R_{k_1,k_2,k_3}$ 
surrounding $(k_1/K,k_2/K)$ as shown in Fig.~\ref{triangles}.

Consider a typical 4-point interaction
%
\be
        V\!\left(\frac{k_1}{K},\frac{k_2}{K},\frac{k_3}{K}\right) =
         \; K \BoxedEPSF{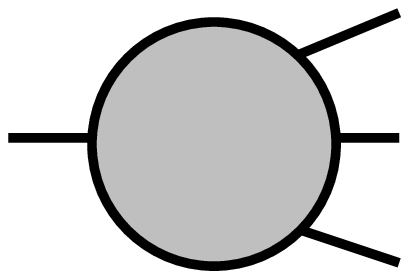 scaled 500}
        \begin{array}{c}k_1\\[10pt] k_2 \\[10pt] k_3
        \end{array} \label{vthree} \; .
\eq
We calculate improved matrix elements associated with
interaction $V(x,y,1-x-y)$ using the expression
\be
  V_{k_1,k_2,k_3} =  \int_0^1 dx \int_0^{1-x} dy\, \phi_{k_1,k_2,k_3}(x,y)
                      V(x,y,1-x-y) \; .
\eq
The integrals are calculated numerically, paying careful attention
to the small momentum regions.

\subsection{Heavy source interactions}

Improvement of the matrix elements for the interactions 
between the heavy sources and the link fields is particularly 
challenging because of the
oscillatory nature of the wavefunction for large 
separation $L$ and the presence of the cutoff $\kmax$.
So far, we have concentrated on removing the leading errors 
which are, as before, associated with small longitudinal momenta
and are of order $1/(\dkmax)^{2\beta}$ and $1/\sqrt{\dkmax}$.  
At this point it makes no sense to include the finite 
$L$ phase factors in the 
improvement scheme because we will also neglect
oscillation of the wavefunction itself in the improved
matrix elements.  The leading error associated with
the improved matrix elements is of order $1/\dkmax$.

For $(\mbox{source-link}) \to (\mbox{source-link})$ 
interactions, we define wavefunctions
\be
    \phi_{k}(x)=  C_k \left\{\begin{array}{cl}
            \displaystyle x^\beta \; ,& 
                k-1/2 < x < k+1/2 \\
           \displaystyle 0 \; , & \mbox{otherwise}
         \end{array} \right.
\eq
where $k,k^\prime\in\{1/2,3/2,\ldots\}$, restricted by
$\dkmax\ge\dklink$, is the DLCQ
momentum of the link field neighboring the heavy source
and $x$ is its continuum longitudinal momentum rescaled by
$\dkmax/(v^+ \sqrt{G^2 N} \kmax)$.  The normalisation is $1=\int dx\, \phi_k(x)^2$.
We calculate improved matrix elements for the 
various interactions using
\be
  V_{k,k^\prime} =  \int dx \, dy\, \phi_{k}(x) \, V(x,y)\,
                      \phi_{k^\prime}(y) 
\eq
with the finite $L$ phase factor replaced by its DLCQ value
\be
        \exp\!\left(\frac{i L \sqrt{G^2 N}(x-y)\kmax}{2 \dkmax}\right)\to
        \exp\!\left(\frac{i L \sqrt{G^2 N}(k-k^\prime)\kmax}{2 \dkmax}\right)
           \; .
\eq

In the case of the mass term for the link field adjacent to
the heavy source, we assign half of the contribution 
to the link-link interactions acting on the interior of the
string, consistent with Eqn.~(\ref{imass}), and the other half to
the heavy-link interaction.  Thus, we use 
\be
        V(x,y)= \frac{\mu^2 \dkmax}{2 x \sqrt{G^2 N} \kmax}\, \delta(x-y)\; .
\eq

For the instantaneous $A_+$ exchange interaction, we must be careful
about how we handle the $1/(x-y)^2$ pole and the 
$\dkmax\ge\dklink$ cutoff.  For the off-diagonal part, we use
\be
  V_{k,k^\prime} = - \frac{\sqrt{G^2 N} \dkmax}{8 \pi (k-k^\prime)^2 \kmax}  
      \exp\!\left(\frac{i L \sqrt{G^2 N}(k-k^\prime)\kmax}{2 \dkmax}\right)
         \int dx \, dy\, 
                \frac{x+y}{\sqrt{x y}}\,\phi_{k}(x)\,
                      \phi_{k^\prime}(y)  \; ,\;\;\;\; k\neq k^\prime\; .
\eq
If the $1/(x-y)^2$ pole were left inside the integral, there
would be a logarithmic divergence when $k=k^\prime\pm 1$.
The diagonal part is obtained by adding and subtracting
the `self-inertia'
\be
  V_{k,k} = \sum_{k^\prime\neq k}^{\dkmax\ge\dklink} 
        \left|V_{k,k^\prime}\right|
                \left(\frac{k^\prime}{k}\right)^\beta
              - \frac{\sqrt{G^2 N} \dkmax}
                {8 \pi\kmax}  \int_0^{K^\prime} dy\, 
                \frac{k+y}{\sqrt{k y}\, (k-y)^2} 
                        \left(\frac{y}{k}\right)^\beta
\eq
where $K^\prime=\dkmax-\dklink+k$.  The integral is solved
numerically with an implicit principle value prescription for the 
pole at $y=k$.  This method for handling
the self-inertias is discussed in Ref.~\cite{mass}.

Turning our attention to the 
$(\mbox{source}) \to (\mbox{source + 2 links})$ interactions,
we proceed in an analogous manner.  Since the wavefunction
becomes constant as any two momenta go to zero, we define  wavefunctions
\be
    \phi_{k_1,k_2}(x,y)=  C_{k_1,k_2} \left\{\begin{array}{cl}
            \displaystyle \frac{(x y)^\beta}{(x+y)^{2 \beta}} \; ,& 
                \begin{array}{l}
                k_1-1/2<x<k_1+1/2\;\; \mbox{and} \\
                k_2-1/2<y<k_2+1/2 \end{array}\\
           \displaystyle 0 \; , & \mbox{otherwise}
         \end{array} \right.
\eq
where $k_1, k_2 \in \{1/2,3/2,\ldots\}$ are the DLCQ momenta
of the two outgoing link fields created adjacent to the source.
The normalisation is $1=\int dx \, dy \, \phi_{k_1,k_2}(x,y)^2$.
We calculate improved matrix elements for the 
various interactions using
\be
  V_{k_1,k_2} =  \int dx \, dy \, \phi_{k_1,k_2}(x,y)\, V(x,y) \; .
\eq
As before, the finite $L$ phase factors in $V(x,y)$ are replaced
by their DLCQ values.

\end{appendix}


\newpage

\end{document}